\documentclass[reprint,superscriptaddress,amssymb,amsmath,aps,prx,longbibliography]{revtex4-2}

\usepackage{minitoc}

\usepackage{graphicx}
\usepackage{dcolumn}
\usepackage{bm}
\usepackage{siunitx}
\usepackage{booktabs}
\usepackage[usenames,dvipsnames]{xcolor}
\usepackage{tcolorbox}
\usepackage{tabularx}
\usepackage{array}
\usepackage{colortbl}
\usepackage{braket}
\usepackage{gensymb}
\usepackage{pdfpages}
\usepackage{amsmath}
\usepackage{mathrsfs}
\usepackage{blindtext}
\usepackage{minitoc}

\makeatletter
\patchcmd{\@outputpage@head}{\@ifx{\LS@rot\@undefined}{}{\LS@rot}}{}{}{}
\makeatother

\tcbuselibrary{skins}

\newcommand{\units}[1]{\,\mathrm{#1}}
\tcbset{tab2/.style={colback=gray!5!white,colframe=black!50!black,colbacktitle=Gray!40!white,
coltitle=black,center title}}

\begin{document}

\title{Microwave Spin Control of a Tin-Vacancy Qubit in Diamond}

\author{Eric I. Rosenthal}
\thanks{ericros@stanford.edu}
\affiliation{E. L. Ginzton Laboratory, Stanford University, Stanford, California 94305, USA}

\author{Christopher P. Anderson}
\affiliation{E. L. Ginzton Laboratory, Stanford University, Stanford, California 94305, USA}

\author{Hannah C. Kleidermacher}
\affiliation{E. L. Ginzton Laboratory, Stanford University, Stanford, California 94305, USA}

\author{Abigail J. Stein}
\affiliation{E. L. Ginzton Laboratory, Stanford University, Stanford, California 94305, USA}

\author{Hope Lee}
\affiliation{E. L. Ginzton Laboratory, Stanford University, Stanford, California 94305, USA}

\author{Jakob Grzesik}
\affiliation{E. L. Ginzton Laboratory, Stanford University, Stanford, California 94305, USA}

\author{Giovanni Scuri}
\affiliation{E. L. Ginzton Laboratory, Stanford University, Stanford, California 94305, USA}

\author{Alison E. Rugar}
\thanks{Present address: SandboxAQ, Palo Alto, California 94305, USA}
\affiliation{E. L. Ginzton Laboratory, Stanford University, Stanford, California 94305, USA}

\author{Daniel Riedel}
\thanks{Present address: AWS Center for Quantum Networking, Boston, Massachusetts, USA}
\affiliation{E. L. Ginzton Laboratory, Stanford University, Stanford, California 94305, USA}

\author{Shahriar Aghaeimeibodi}
\thanks{Present address: AWS Center for Quantum Computing, San Francisco, California, USA}
\affiliation{E. L. Ginzton Laboratory, Stanford University, Stanford, California 94305, USA}

\author{Geun Ho Ahn}
\affiliation{E. L. Ginzton Laboratory, Stanford University, Stanford, California 94305, USA}

\author{Kasper Van Gasse}
\affiliation{E. L. Ginzton Laboratory, Stanford University, Stanford, California 94305, USA}

\author{Jelena Vu\v{c}kovi\'{c}}
\affiliation{E. L. Ginzton Laboratory, Stanford University, Stanford, California 94305, USA}

\date{\today}
\begin{abstract}
    The negatively charged tin-vacancy (SnV$^-$) center in diamond is a promising solid-state qubit for applications in quantum networking due to its high quantum efficiency, strong zero phonon emission, and reduced sensitivity to electrical noise. The SnV$^-$ has a large spin-orbit coupling, which allows for long spin lifetimes at elevated temperatures, but unfortunately suppresses the magnetic dipole transitions desired for quantum control. Here, by use of a naturally strained center, we overcome this limitation and achieve high-fidelity microwave spin control. We demonstrate a $\pi$-pulse fidelity of up to $99.51 \pm 0.03\%$ and a Hahn-echo coherence time of $T_2^\mathrm{echo} = 170.0 \pm 2.8 \units{\mu s}$, both the highest yet reported for SnV$^-$ platform. This performance comes without compromise to optical stability, and is demonstrated at 1.7 K where ample cooling power is available to mitigate drive-induced heating. These results pave the way for SnV$^-$ spins to be used as a building block for future quantum technologies.
\end{abstract}

\maketitle

\section{Introduction}

Networked entanglement between spatially separated nodes promises to revolutionize quantum computing, sensing and communication \cite{kimble:2008}. Solid-state quantum emitters hold potential as the building blocks of such networks \cite{wehner:2018}. These nodes require many features, including efficient photon collection, spin-photon entanglement, single-shot readout, and the coherent manipulation of long-lived spins \cite{wolfowicz:2021}.

The most sophisticated quantum networks today use the nitrogen-vacancy (NV$^-$) center in diamond \cite{pompili:2021,hermans:2022}, but are limited by its properties including low emission into its zero-phonon line and noisy optical transitions. In comparison, group IV centers in diamond have many advantages as qubits to use for the next generation of quantum networks. These defects, the silicon (SiV$^-$), germanium (GeV$^-$), tin (SnV$^-$), and lead (PbV$^-$) vacancy centers, all have strong zero-phonon line emission and an inversion symmetric structure, reducing their sensitivity to electrical noise \cite{thiering:2018}. This insensitivity is crucial because it enables stable, narrow emission within photonic nanostructures \cite{rugar:2021,martinez:2022,parker:2023}, serving as an efficient spin-photon interface key for quantum networks. Of the group IV color centers, the SiV$^-$ is the most technologically mature to date, with demonstrations of memory enhanced quantum communication \cite{bhaskar:2020} and integrated error detection \cite{stas:2022}.

The SiV$^-$ has the smallest spin-orbit coupling among this family of defect centers, with a strain-free ground state splitting of $\approx 50 \units{GHz}$. This makes it necessary to operate the SiV$^-$ in a dilution refrigerator (temperature $\lesssim 100 \units{mK}$) to avoid decoherence due to phonons \cite{jahnke:2015,pingault:2017,sohn:2018,meesala:2018,nguyen:2019}. However, the limited cooling power at this temperature makes it challenging to implement complex pulse sequences due to drive-induced heating \cite{bhaskar:2020}. 

\begin{figure}[htb!] 
\begin{center}
\includegraphics[width=1.0\columnwidth]{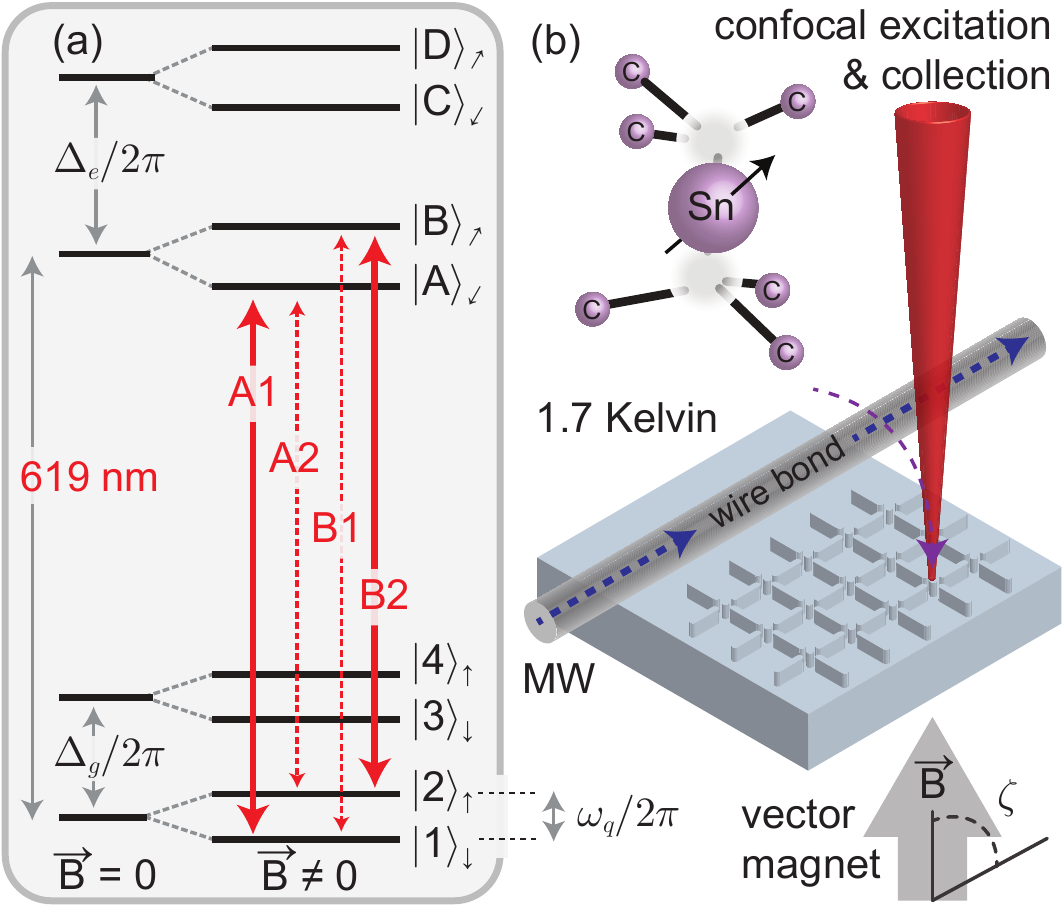}
\caption {
(a) Energy diagram of the negatively charged tin-vacancy (SnV$^-$) center in diamond. The ground and excited state manifolds are split into two pairs of states, $\ket{1}$, $\ket{2}$ and $\ket{3}$, $\ket{4}$, which are separated by the ground state splitting $\Delta_g/2\pi = 903 \units{GHz}$ for this SnV$^-$ (Fig.~\ref{fig:pl}a). Pairs $\ket{A}, \ket{B}$ and $\ket{C}, \ket{D}$ are separated by the excited state splitting $\Delta_e/2\pi \approx 3000 \units{GHz}$. A magnetic field further splits each pair via the Zeeman effect, creating distinct transitions: A1 and B2 (spin preserving) and A2 and B1 (spin flipping). States $\ket{1}$ and $\ket{2}$ are split by the qubit frequency $\omega_q/2\pi$. (b) In this work, an SnV$^-$ within a nanopillar is cooled to 1.7 K, and excited and measured confocally. The qubit is controlled by microwave (MW) pulses, delivered via a wire bond draped so that its center is $\approx 60 \units{\mu m}$ from the SnV$^-$. The diamond is oriented with $\langle100\rangle$ along the Z axis of the vector magnet.
}
\label{fig:intro}
\end{center}
\end{figure}

On the other hand, the strain-free ground state splitting of the SnV$^-$ is $\approx850 \units{GHz}$ \cite{iwasaki:2017}; see Fig.~\ref{fig:intro}a. This larger energy allows for coherent operation at higher temperatures above 1 K, where exponentially more cooling power is available \cite{pobell:2007,wang:2014} and cryostat technology poses fewer challenges to scaling. In fact, recent advances in the SiV$^-$ simply use highly strained emitters, where the ground state splitting approaches that of the SnV$^-$ (e.g., 554 GHz in Ref.~\cite{stas:2022}).

Recent experimental progress using the SnV$^-$ platform includes characterization of its spin and optical properties \cite{tchernij:2017,iwasaki:2017,rugar:2019,trusheim:2020,gorlitz:2020} including large hyperfine interactions \cite{parker:2023,harris:2023}, nanophotonic integration \cite{rugar:2020,rugar:2021}, Stark tuning \cite{desantis:2021,aghaeimeibodi:2021}, spin control using optical Raman driving \cite{debroux:2021}, and single-shot nuclear spin readout \cite{parker:2023}. However, the SnV$^-$'s $\gtrsim 30 \units{MHz}$ transform limited optical linewidth \cite{iwasaki:2017,trusheim:2020,gorlitz:2020} presents a challenge to high-fidelity optical spin control. To prevent drive-induced dephasing, the optical control pulse must be strong and detuned by many linewidths (see Appendix~\ref{sec:optical_spin_control}). For example, this drive-induced dephasing limits the achievable gate fidelity in Ref.~\cite{debroux:2021} such that coherence is lost after several pulses. This inhibits dynamical decoupling schemes desired to extend coherence, and the multipulse control needed to utilize long-lived nuclear registers. Because gate fidelity must be improved for SnV$^-$'s to have future use in quantum technology, alternate control techniques should be explored.

Here, we overcome this challenge by instead using microwave driving to demonstrate high-fidelity coherent control of a single SnV$^-$ spin qubit. Because of mixed spin and orbital character, the qubit's transition is forbidden to first order for an unstrained emitter \cite{trusheim:2020,debroux:2021}. However, strain perturbs the SnV$^-$ to allow for direct driving with microwaves without compromising its stable, narrow optical lines \cite{sukachev:2017,pingault:2017,debroux:2021}. First, we characterize the SnV$^-$ level structure as a function of magnetic field. This gives a precise measurement of the SnV$^-$ Hamiltonian including strain, and illuminates favorable operating conditions for microwave spin control. We then demonstrate spin control and characterize its fidelity. Using high-fidelity gates we show the qubit's coherence time can be extended to hundreds of microseconds using dynamical decoupling. Finally, we use spin control to understand sources of noise affecting the qubit. By studying coherence time as a function of dynamical decoupling sequence, temperature, and magnetic field, we determine straightforward ways in which future SnV$^-$ experiments may be improved. Our work allows for exploration of the rich spin physics of these systems and enables their use as a future building block for quantum networks. 

\begin{figure*}[tb!] 
\begin{center}
\includegraphics[width=1.98\columnwidth]{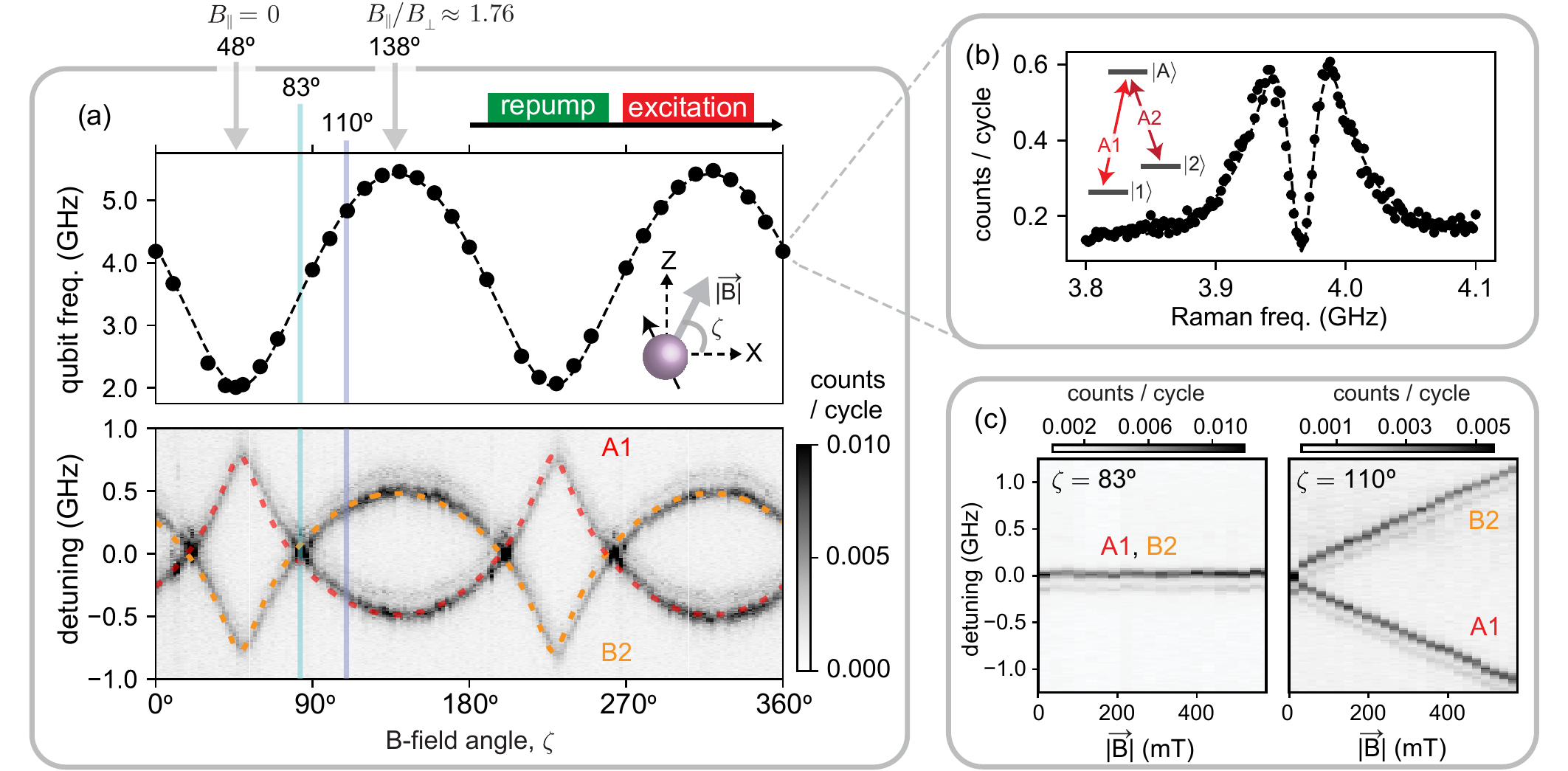}
\caption{
SnV$^-$ level structure. (a) Transition frequencies versus angle $\zeta$ of the magnetic field $\Vec{B}$ in the lab coordinates, such that $B_\mathrm{X}=|\Vec{B}|\cos(\zeta)$ and $B_\mathrm{Z}=|\Vec{B}|\sin(\zeta)$. Top: qubit frequency $\omega_q/2\pi$ (splitting between $\ket{1}$ \& $\ket{2}$ states), measured using coherent population trapping (CPT). Top and bottom panels share an $x$ axis. Bottom: spin preserving optical transitions (A1 and B2), measured using photoluminescence excitation (PLE). Dashed lines in both the top and bottom panels are a fit obtained from the SnV$^-$ Hamiltonian (Eq.~\ref{eqn:HSnV}), using the parameters in Table~\ref{tab:SnVparameters}. At $\zeta=48^{\degree}$, $\Vec{B}$ is orthogonal to the spin's dipole, such that its parallel component is $B_\parallel=0$ and its perpendicular component $B_\perp$ is nonzero. At $\zeta=138^{\degree}$, $\Vec{B}$ is closest to aligned with the spin dipole, which for the configuration in this experiment results in $B_\parallel/B_\perp \approx 1.76$. (b) Representative example of CPT. Dashed line is a numerical model. (c) Zeeman effect: PLE versus $|\Vec{B}|$ at $\zeta = 83^{\degree}$ (left) and $\zeta = 110^{\degree}$ (right). The data in panels (a) and (c) are taken at 4 K, for increased signal due to reduced spin initialization under resonant excitation. The faint secondary lines following each transition are due to multimode behavior of the excitation laser.
}
\label{fig:ple}
\end{center}
\end{figure*}

\section{Strain considerations for S\MakeLowercase{n}V$^-$ centers}
Group IV centers in diamond present a conundrum: symmetry protects from noise but also inhibits direct control of the spin transition. To understand microwave spin control, it is of crucial importance to quantify the degree to which symmetry is broken via strain. This will enable us to understand why microwave spin control works and to analyze associated trade-offs.

To that aim, we first characterize the level structure and Hamiltonian of the SnV$^-$ center used in this work. We use a center embedded within a diamond nanopillar in order to increase light collection efficiency, Fig.~\ref{fig:intro}b. The center is optically excited and read out confocally. A g$^\mathrm{(2)}$ correlation measurement confirms this is a single emitter, Fig.~\ref{fig:pl}a.

To estimate ground state strain $\Upsilon_g$, first we measure the photoluminescence (PL) spectrum of a single SnV$^-$ under excitation with above-resonant light (532 nm) at zero magnetic field, Fig.~\ref{fig:pl}b. Two bright transitions are measured near 619 nm, which are split by the ground state splitting $\Delta_g/2\pi = 903.0\pm 0.7 \units{GHz}$, Fig.~\ref{fig:intro}a. This is the difference in frequency between the two lowest pairs of levels in the ground state manifold, and is related to both spin-orbit coupling $\lambda_g$ and strain $\Upsilon_g$ by $\Delta_g = \sqrt{\lambda_g^2 + 4\Upsilon_g^2}$.

Our measured $\Delta_g/2\pi$ is larger than reported in previous SnV$^-$ work (between 820 and 850 GHz in Refs.~\cite{iwasaki:2017,rugar:2019,trusheim:2020,gorlitz:2020,gorlitz:2022}), indicating greater strain. However, due to a lack of precise knowledge of $\lambda_g$, the relative contributions of spin-orbit coupling and strain cannot be determined from a PL measurement alone. To determine both, we fix magnetic field amplitude at $|\Vec{B}| = 184 \units{mT}$ and sweep its angular orientation along a circle formed by the X and Z coils of our vector magnet (see Fig.~\ref{fig:coordinates}a for a diagram of this coordinate system). While doing so we measure the SnV$^-$'s qubit and optical transitions as a function of field angle, Fig.~\ref{fig:ple}a. Qubit transitions (the frequency difference $\omega_q/2\pi$ between the $\ket{1}$ and $\ket{2}$ states) are measured using coherent population trapping (CPT), Fig.~\ref{fig:ple}b. Here, CPT is a useful tool to determine $\omega_q/2\pi$ as a precursor to calibrating microwave spin control. Finally, to complete characterization of the SnV$^-$ level structure we show that the frequency difference between the spin preserving A1 and B2 transitions splits linearly with magnetic field, Fig.~\ref{fig:ple}c. The rate of splitting is highly dependent on field orientation, due to a competition between spin-orbit and the Zeeman effect.

This combined dataset of spin and optical transition frequencies is fit to a model Hamiltonian (Eq.~\ref{eqn:HSnV}) in order to measure the parameters of this SnV$^-$. We determine that this system displays ``moderate'' strain: with a ground state strain of $\Upsilon_g/2\pi = 177.67 \pm 1.37 \units{GHz}$, compared to a ground state spin-orbit coupling of $\lambda_g/2\pi = 830.15 \pm 1.42 \units{GHz}$. This value of strain is somewhat larger than other values reported in the literature, e.g. 80 GHz in Ref.~\cite{trusheim:2020} and 95 GHz in Ref.~\cite{debroux:2021}, but remains in the limit where spin-orbit coupling is dominant. 

Simulations of the SnV$^-$ Hamiltonian with this level of strain predict direct magnetic driving of the qubit transition to be weakly allowed (see Appendix~\ref{sec:simulated_behavior}). These simulations predict a Rabi rate $\Omega_\mathrm{MW}/2\pi$ of between 3 -- 22 MHz for a microwave drive field of $|\Vec{b}| = 1.6 \units{mT}$ at the SnV$^-$ location. Variation within this range is dependent on the orientation of the static and bias magnetic fields compared to the spin dipole moment. This drive field corresponds to 0.5 A of microwave current traveling through a bias line $60 \units{\mu m}$ from the qubit, reflecting the parameters used in this work. We note that with this configuration, the Rabi rate for a free-electron spin with a fully allowed magnetic dipole transition would be $\approx$22 MHz, such that the suppression due to the spin-orbit interaction is moderate. This implies that coherent control of our SnV$^-$ should require no more than 10 times more microwave power than an NV$^-$ center in diamond, for example.

Despite enough strain to enable high-fidelity microwave spin control, the SnV$^-$'s optical transitions remain stable with a linewidth of $60 \pm 10 \units{MHz}$, Fig.~\ref{fig:stability}, close to their transform limit of $30 \units{MHz}$ \cite{iwasaki:2017,trusheim:2020,gorlitz:2020}. This is expected: the Jahn-Teller Hamiltonian reduces the symmetry of group IV centers from point group $D_{3d}$ (unstrained) to point group $C_{2h}$ (strained), while maintaining inversion symmetry (point group $C_i$) as an irreproducible representation \cite{hepp:2014,thiering:2018}. In other words, uniaxial strain can deform the SnV$^-$ wave function and shift orbital states, but does not break the inversion symmetry along the defect axis. Thus, a strained emitter enables microwave spin control but keeps the optical transitions first order insensitive to electrical noise. Unfortunately, strain does reduce the cyclicity of the spin preserving transitions A1 and B2, to the detriment of spin selective readout (see Appendix~\ref{sec:simulated_behavior}). However, regimes that balance this trade-off have been found in SiV$^-$ centers \cite{sukachev:2017}, and can hold as well for the SnV$^-$.

\section{Spin Control}

\subsection{Demonstration}

\begin{figure*}[tb!] 
\begin{center}
\includegraphics[width=1.98\columnwidth]{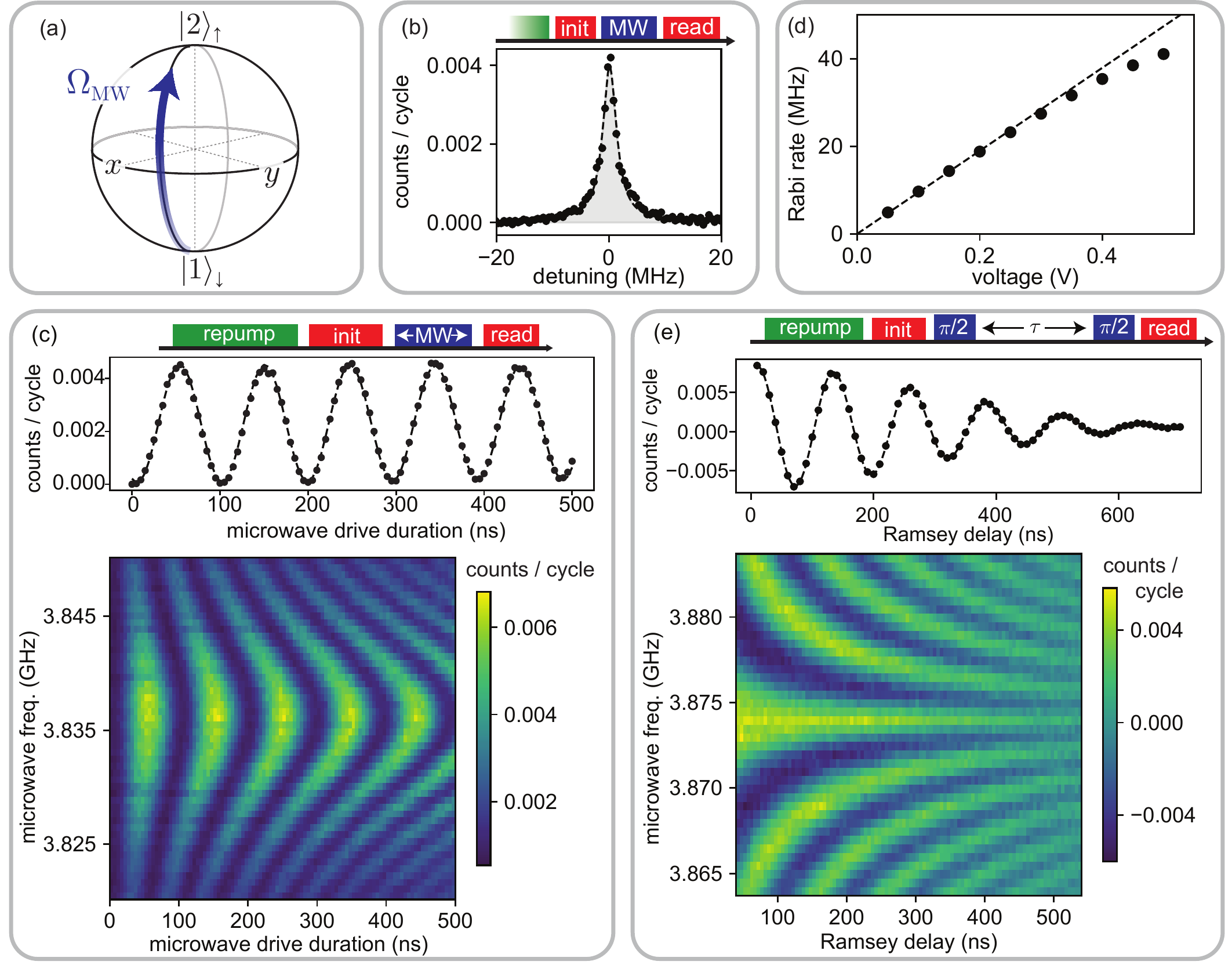}
\caption{
Microwave spin control. (a) The SnV$^-$'s lowest two states, $\ket{1}$ and $\ket{2}$, are controlled at Rabi rate $\Omega_\mathrm{MW}$ using a coherent microwave drive. (b) Pulsed optically detected magnetic resonance (ODMR). Dashed line is a Lorentzian fit with linewidth $3.09 \pm 0.06 \units{MHz}$ and center frequency 3.7912 GHz. 
(c) Rabi oscillations at rate $\Omega_\mathrm{MW}/2\pi = 20.67 \pm 0.02 \units{MHz}$ (a 48.4 ns $\pi$ pulse) using $48 \units{dBm}$ of power into the cryostat. Data in (b) and (c) are the difference in signal (counts per cycle) between experimental cycles with the microwave drive turned on or off. (d) Rabi rate is linear with microwave amplitude, proportional to the voltage applied to an IQ (in-phase and quadrature) mixer. Deviation from linear behavior at high voltage is due to saturation of the mixer. (e) Ramsey measurement. Data are modeled by a sinusoid with decaying envelope $e^{-(t/T_2^*)^\xi}$, where we fit $T_2^* = 396.6 \pm 2.29 \units{ns}$ and stretching exponent $\xi=2.077 \pm 0.036$. Data are the difference in signal (counts per cycle) between experiments that use a pulse sequence $[(\pi/2)$-$\tau$-$(\pi/2)]$, versus a sequence of $[(\pi/2)$-$\tau$-$(-\pi/2)]$. 
}
\label{fig:rabi}
\end{center}
\end{figure*}

With a clearer picture of this SnV$^-$ Hamiltonian under strain, we use a microwave field to coherently control the qubit formed by its two lowest states, $\ket{1}$ and $\ket{2}$; see Fig.~\ref{fig:rabi}a. 

For data in this section, we operate at a magnetic field of $|\Vec{B}| = 150 \units{mT}$, oriented at the angle $\zeta=110^{\degree}$ (purple vertical line in Fig.~\ref{fig:ple}a). Here the qubit states are split by $\approx 3.8 \units{GHz}$. The spin preserving A1 and B2 transitions are closer together, split by $\approx 0.5 \units{GHz}$, but are still spectrally resolved compared to their $\approx 60 \units{MHz}$ linewidth. These transitions are selectively driven to enable initialization and readout of the qubit state. For example, driving on the B2 transition initializes the qubit in $\ket{1}$, Fig.~\ref{fig:init}. To control the qubit's state, we run microwave current through an aluminum wire bond draped across the chip, $\approx 60 \units{\mu m}$ away from the nanopillar at the closest point. 

We first characterize our qubit using optically detected magnetic resonance (ODMR), Fig.~\ref{fig:rabi}b, in which a microwave pulse of variable frequency is applied between initialization and readout pulses. When sweeping the frequency of this microwave pulse, we measure a peak at 3.7912 GHz (linewidth of $3.09 \pm 0.06 \units{MHz}$, when fitting a Lorentzian). Fixing at this frequency and sweeping the microwave pulse duration yields coherent rotation of the qubit state around the Bloch sphere, i.e. Rabi oscillations. An example of this is shown in Fig.~\ref{fig:rabi}c at a Rabi rate of $\Omega_\mathrm{MW}/2\pi = 20.67 \pm 0.02 \units{MHz}$ corresponding to a $\pi$ rotation in $48.4 \units{ns}$. High-fidelity control is illustrated by the preservation of readout contrast over many oscillations.

The Rabi rate is linearly proportional to microwave drive amplitude, Fig.~\ref{fig:rabi}d, which we characterize up to $\approx40 \units{MHz}$ (25 ns $\pi$ pulses). Faster manipulation is limited by the bandwidth of the control electronics used in this experiment. 
At higher Rabi rates contrast reduces after fewer oscillations, Fig.~\ref{fig:rabi_vs_power}, which we ascribe to drive-induced heating (see Appendix~\ref{sec:infidelity}). In conclusion, the microwave Rabi rate can be as fast as desired, albeit requiring high microwave power and with a trade-off between speed and heating induced infidelity.

Next, we measure the qubit's dephasing time using Ramsey interferometry, shown in Fig.~\ref{fig:rabi}e. Fitting decay of the measured Ramsey fringes as a function of time demonstrates that the qubit's superposition state is maintained for a characteristic timescale of $T_2^* = 396.6 \pm 2.29 \units{ns}$ and a stretching exponent of $\xi=2.077 \pm 0.036$. This time is similar to other measurements of group IV centers that have been implanted into natural isotopic abundance diamond, which range from hundreds of nanoseconds to several microseconds \cite{sukachev:2017,bauch:2020,debroux:2021}. 

\subsection{Gate characterization}

\begin{figure}[tb!] 
\begin{center}
\includegraphics[width=1.0\columnwidth]{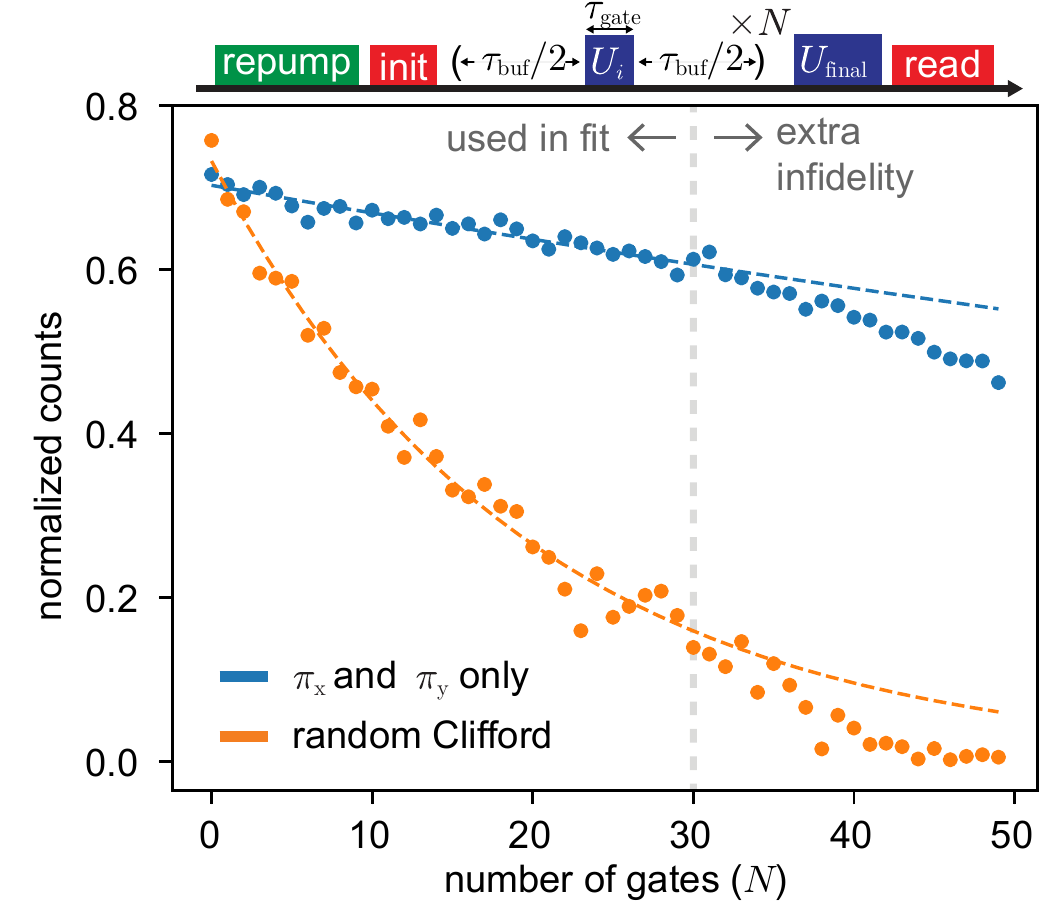}
\caption{
Gate characterization. $N$ gates are applied to the qubit, each of duration $\tau_\mathrm{gate}$ separated by a buffer $\tau_\mathrm{buf}$ during which the microwave drive is off. Gates are either $\pi$ rotations of $\tau_\mathrm{gate} = 54 \units{ns}$ chosen at random from $U_i \in \{\pi_x, \pi_y\}$ and with $\tau_\mathrm{buf} = 480 \units{ns}$ or randomized benchmarking \cite{magesan:2011}: a random Clifford gate, $U_i \in \{I, \pi_x, \pi_y, \pm \pi_x/2, \pm \pi_y/2\}$, with $\tau_\mathrm{gate} = 52 \units{ns}$ for $\pi$ pulses, $\tau_\mathrm{gate} = 26 \units{ns}$ for $\pi/2$ pulses, and the buffer time chosen such that $\tau_\mathrm{gate} + \tau_\mathrm{buf} = 92 \units{ns}$ regardless of the type of gate. Randomized benchmarking data are averaged over 17 different random sequences at each value of $N$. After $N$ operations, a final $(N+1)$th operation $U_\mathrm{final}$ is applied to project the qubit into the $\ket{1}$ state or, in an alternating experimental cycle, to the $\ket{2}$ state. The $y$ axis is the difference in signal (number of detected photons per cycle) between these experiments, divided by the sum of signal in both. Data up to $N=30$ are modeled by the function $a\mathcal{F}^N$ where $\mathcal{F}$ is the gate fidelity, fit to $99.51\% \pm 0.03\%$ for a $\pi$ pulse and $95.04\% \pm 0.14\%$ for a random Clifford. At greater $N$, fidelity worsens more quickly with $N$ than this model. This extra infidelity may be due to heating effects.
}
\label{fig:rb}
\end{center}
\end{figure}

To quantify spin control fidelity, we now sweep the number of gate operations, $N$, each denoted in Fig.~\ref{fig:rb} by the unitary rotation $U_i$. This procedure is applied either using only $\pi$ pulses such that $U_i \in \{\pi_x, \pi_y\}$, or using randomized benchmarking \cite{magesan:2011}: $U_i \in \{I, \pi_x, \pi_y, \pm \pi_x/2, \pm \pi_y/2\}$.

Data from the first $N\leq30$ gates in Fig.~\ref{fig:rb} are fit to the function $a \mathcal{F}^N$. This yields a $\pi$-pulse fidelity of $\mathcal{F} = 99.51\% \pm 0.03\%$ using $54 \units{ns}$ pulses, and an average random Clifford fidelity of $\mathcal{F} = 95.04\% \pm 0.14\%$ using $52 \units{ns}$ pulses. The Clifford fidelity is lower than the $\pi$-pulse fidelity because for this qubit, $T_2^* \ll T_1$, such that when the qubit is prepared in a superposition state (commonly occurring in the randomized benchmarking experiment but not with successive $\pi$ pulses), the dominant source of error is dephasing. Slower pulses than those used in Fig.~\ref{fig:rb} result in lower fidelity, as they are increasingly susceptible to dephasing errors as pulse time approaches $T_2^*$. In our experiment, faster pulses than those used in Fig.~\ref{fig:rb} are limited by the bandwidth of the pulse generation electronics and by drive-induced heating.

Extra infidelity occurs as $N$ increases beyond approximately 30 pulses, likely due to drive-induced heating for this particular experiment (see Appendix~\ref{sec:infidelity}). While the base temperature of the cryostat does not rise above 1.8 K during the measurements in Fig.~\ref{fig:rb} (from a base temperature of 1.7 K), local and instantaneous heating can be greater than the sample thermometer would suggest. 

Gate fidelity can be improved. Infidelity may be reduced by lengthening the delay time between waveform sequences, at the expense of a slower experimental repetition rate. For the sequence of $\pi$ pulses only, lengthening $\tau_\mathrm{buf}$ also improves fidelity. In future devices, coherence time $T_2^*$ can be improved by switching to isotopically pure diamond \cite{sukachev:2017}, improving Clifford fidelity. Heating can also be mitigated by lowering the required microwave bias current. This can be done by using a SnV$^-$ center with greater natural strain, engineering the microwave drive current to be closer to the qubit, or optimizing the orientation of the static magnetic field and microwave bias field relative to the emitter's spin axis (see Fig.~\ref{fig:sims_vs_strain}).

\subsection{Coherence}
\subsubsection{Dynamical decoupling}

\begin{figure}[tb!] 
\begin{center}
\includegraphics[width=1.0\columnwidth]{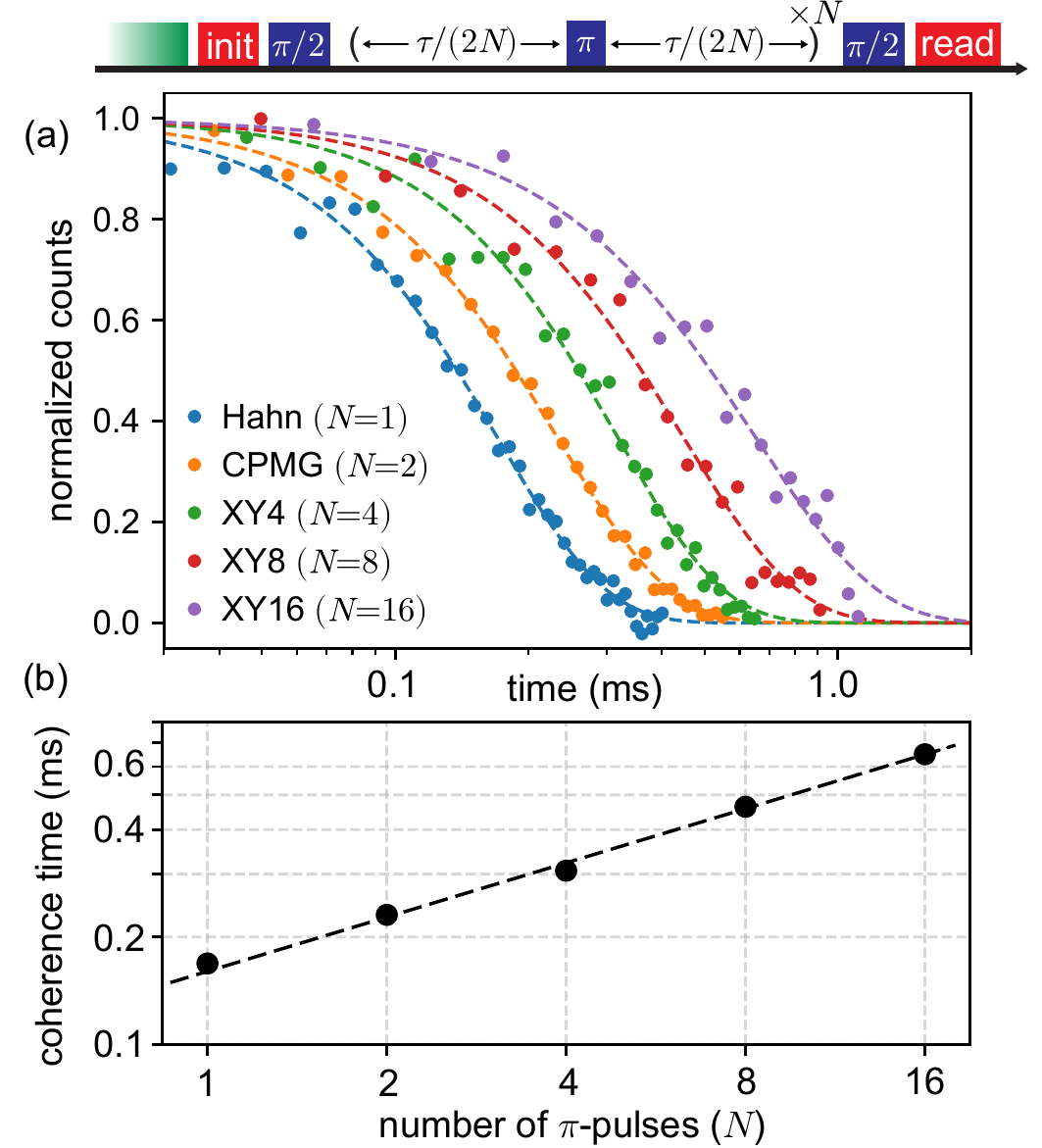}
\caption{
Coherence time. (a) Coherence time is measured using dynamical decoupling. Data are fit to the decaying envelope $e^{-(t/T_2)^\xi}$. The measured time is $T_2^\mathrm{echo} = 170.0 \pm 2.8 \units{\mu s}$ using a Hahn-echo sequence ($N=1$) \cite{hahn:1950}, and up to $T_2^\mathrm{XY16} = 650.2 \pm 28.4 \units{\mu s}$ using $N=16$ decoupling pulses. Measurements with $N>1$ use a Carr-Purcell-Meiboom-Gill (CPMG) sequence \cite{carr:1954,meiboom:1958}, and measurements with $N=4,8,16$ use an XY style variant \cite{gullion:1990}. The $y$ axis is the difference in signal (counts per cycle) between experimental cycles that use a pulse sequence $[(\pi/2)$-(decoupling pulses)-$(\pi/2)]$ versus a sequence of $[(\pi/2)$-(decoupling pulses)-$(-\pi/2)]$, divided by the sum of signal in both. (b) $T_2$ versus the number of decoupling pulses $N$. The stretch exponent $\xi$ is between $1.6$ and $1.8$ for all fits, with standard deviations of up to $\pm0.17$. Dashed line is the model $a N^\chi$, with $\chi=0.505\pm0.016$. Data are plotted on a log-log scale and error bars are within the data points.
}
\label{fig:t2}
\end{center}
\end{figure}

Using these high-fidelity gates, we now measure the qubit's coherence time using dynamical decoupling, Fig.~\ref{fig:t2}a. This extends coherence by orders of magnitude; for example we measure a Hahn-echo coherence time of $T_2^\mathrm{echo} = 170.0 \pm 2.8 \units{\mu s}$ \cite{hahn:1950}. We also increase the number of decoupling pulses, and measure up to $T_2^\mathrm{XY16} = 650 \pm 28 \units{\mu s}$ using an XY16 sequence \cite{carr:1954,meiboom:1958,gullion:1990}. Next we study coherence time as a function of the number of decoupling pulses $N$, Fig.~\ref{fig:t2}b. We model coherence time as proportional to $a N^\chi$ and fit to obtain $\chi=0.505\pm0.016$. This $\approx\sqrt{N}$ scaling is consistent with a $1/f$ noise source dominating decoherence \cite{medford:2012,bargill:2012}. 

Our measured coherence times are roughly in line with the variability present in previous works with group IV centers \cite{sukachev:2017,debroux:2021}. However, compared to the previous SnV$^-$ measurement in Ref.~\cite{debroux:2021} our $T_2^*$ time ($\approx$400 ns, Fig.~\ref{fig:rabi}e) is approximately a factor of 3 shorter, and our $T_2^\mathrm{echo}$ time ($\approx170 \units{\mu s}$, Fig.~\ref{fig:t2}) is approximately 6 times longer. 

To further understand sources of decoherence, we use the cluster-correlation-expansion (CCE) technique \cite{onizhuk:2021} to numerically simulate dephasing from both the local nuclear and electronic spin bath (see Appendix~\ref{sec:coherence_model}). From these calculations, we find our measured $T_2^*$ falls near the range expected from simulation of a nuclear spin bath in naturally abundant diamond (1.1\% $^{13}$C). 

However, our measured $T_2^\mathrm{echo}$ is significantly shorter than expected from the nuclear spin bath alone, which is predicted to be $T_2^\mathrm{echo} \approx 800 \pm 200 \units{\mu s}$, Fig.~\ref{fig:spin_bath}. We therefore attribute the extra measured dephasing to a bath of electron-spin states in the diamond with a concentration of $\approx8\times10^{16} \units{cm}^3$, likely arising from other Sn impurities and $S=1/2$ vacancy centers created during implantation (see Appendix~\ref{sec:coherence_model}). To increase $T_2^\mathrm{echo}$, the concentration of nearby electron spins may be reduced by changing the sample's implantation and annealing conditions; for example by using shallow ion implantation and overgrowth (SIIG) \cite{rugar:2020b}.

\subsubsection{Coherence versus magnetic field}

\begin{figure}[tb!] 
\begin{center}
\includegraphics[width=1.0\columnwidth]{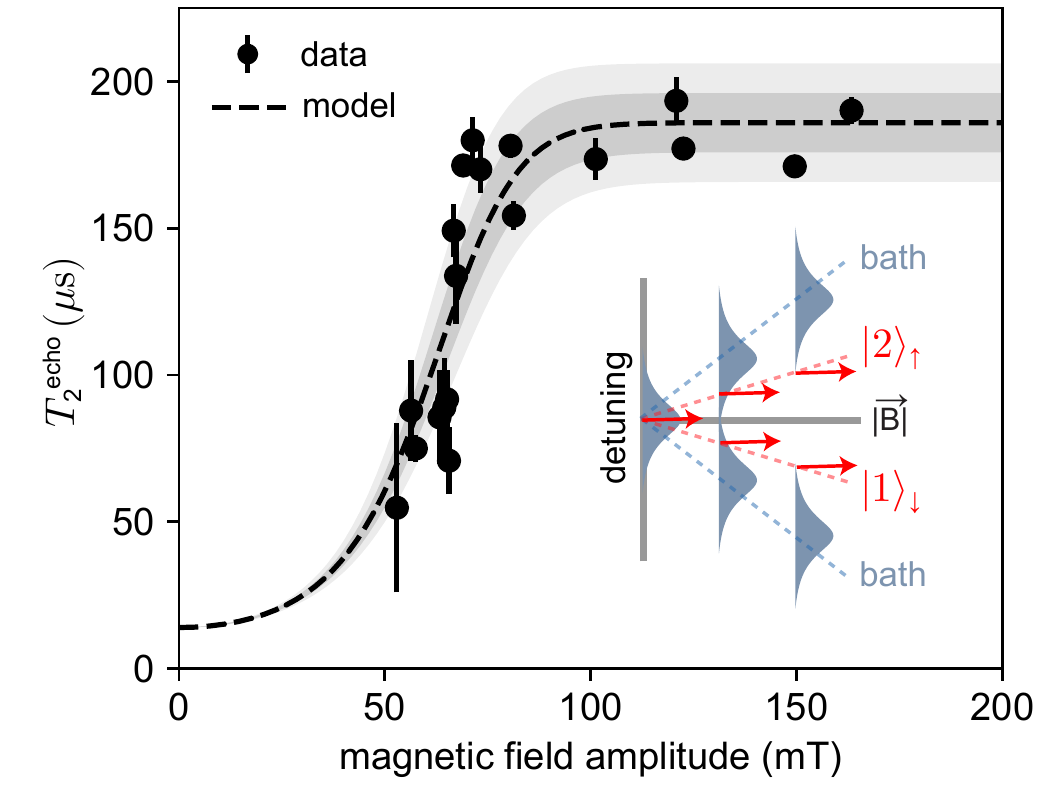}
\caption{
Coherence time $T_2^\mathrm{echo}$ versus magnetic field amplitude, for a field oriented at $\zeta=110^{\degree}$. The predicted increase then plateau is expected from the SnV$^-$ qubit detuning from the predominantly $S=1/2$ bath, due to the anisotropy in its electronic $g$ factor (inset). The black dashed line is a semiclassical model using the qubit's $g$ factor and its maximum $T_2^\mathrm{echo}$ time as free parameters. Dark (light) shaded regions indicate the range over which both fit parameters are changed by $\pm1$ ($\pm2$) standard deviations, respectively.
}
\label{fig:t2_vs_b}
\end{center}
\end{figure}

One complication of $S=1/2$ qubits such as the group IV centers in diamond is that the majority of undesirable and noisy bath spins are also $S=1/2$ with a $g$ factor close to 2 \cite{brosious:1974}. Therefore, control pulses on the central spin will also manipulate the uncontrolled electron spins of the bath. In this case, refocusing of the magnetic noise from these spins is ineffective, referred to as instantaneous diffusion \cite{wolfowicz:2021}.

However, the Hamiltonian of the SnV$^-$ (in particular, the spin-orbit interaction) renders its spin anisotropic with a $g$ factor that can differ from the $g=2$ electron-spin states of the bath \cite{thiering:2018}, Fig.~\ref{fig:ple}a. As a result, at large enough fields the SnV$^-$ qubit states separate from the bath, such that control pulses do not flip the spins in the environment.

To understand this limitation to coherence time, we measure $T_2^\mathrm{echo}$ as a function of magnetic field amplitude, Fig.~\ref{fig:t2_vs_b}. We find that at low field $T_2^\mathrm{echo}$ is significantly reduced, but then rises with field amplitude and saturates at $\gtrsim 100 \units{mT}$. This is understood as reaching the regime where the control pulse bandwidth is much less than the detuning between the qubit and the bath.

We reproduce this feature with a simple model. Using our $\pi$-pulse time of $\approx 50 \units{ns}$ and a simulated electron-spin bath concentration of $8\times10^{16}$ cm$^{-3}$ (chosen to model our measured $T_2^\mathrm{echo}$), we can estimate the fraction of the bath that contributes to instantaneous diffusion and therefore compute the decoherence rate with an approximate semiclassical model \cite{wolfowicz:2021} (see Appendix~\ref{sec:coherence_model}). We fit the data in Fig. \ref{fig:t2_vs_b} to this model, with the only free parameters being the effective $g$ factor and $T_2^\mathrm{echo}$ in the high field limit. This fit returns $g=1.873\pm0.004$. From the anisotropic and angle dependent qubit frequency, Fig.~\ref{fig:ple}a, we estimate that for this experiment, $g\approx1.86$, consistent with this fit. We therefore conclude that dynamical decoupling will extend the coherence time of SnV$^-$ qubits, so long as they are operated in a regime not dominated by instantaneous diffusion. This regime can be achieved by operating at high enough magnetic fields at appropriate angle.

\subsubsection{Coherence versus temperature}
\begin{figure}[tb!] 
\begin{center}
\includegraphics[width=1.0\columnwidth]{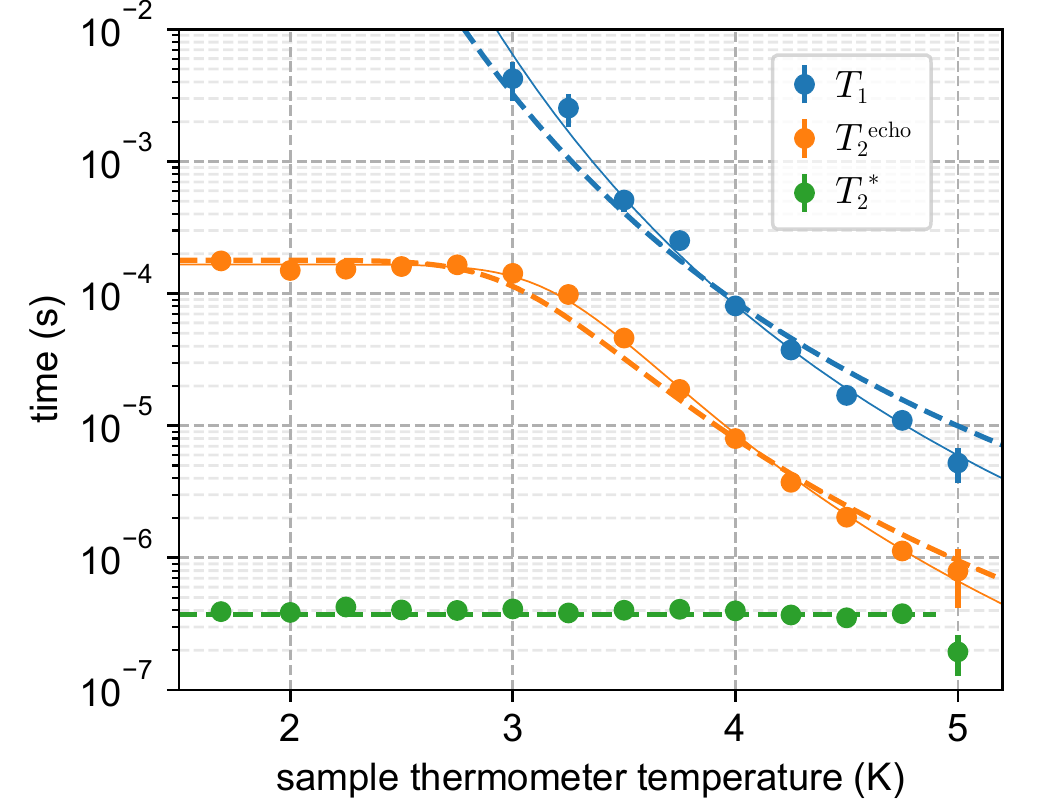}
\caption{
Temperature dependence of $T_1$, $T_2^\mathrm{echo}$, and $T_2^*$. The blue and orange lines are fits to Eq.~\ref{eqn:T1_vs_temp_model} for $T_1$ and Eq.~\ref{eqn:T2_vs_temp_model} for $T_2^\mathrm{echo}$, respectively. Green dashed line is constant at $392 \units{ns}$, the mean value of $T_2^* $ data up to 4.75 K. Thin solid lines are phenomenological models in which a scaling factor $\alpha$ is included to multiply the term $\Delta_g/T$. This modifies the denominator of Eq.~\ref{eqn:T1_vs_temp_model} to $e^{\alpha(\hbar\Delta_g/k_B T)}-1$. We obtain $\alpha=1.207\pm0.045$ by fitting to $T_1$ data and $\alpha=1.208\pm0.029$ by fitting to $T_2^\mathrm{echo}$ data.
}
\label{fig:t2_vs_temp}
\end{center}
\end{figure}

Qubit control data have been taken thus far at $\approx 1.7 \units{K}$. To characterize the temperature at which the SnV$^-$ qubit is practical to use, we now measure $T_1$, $T_2^*$, and $T_2^\mathrm{echo}$ as a function of temperature up to 5 K, Fig.~\ref{fig:t2_vs_temp}.

We measure a qubit energy relaxation time of $T_1 = 4.23 \pm 1.37 \units{ms}$ at 3 K, which decreases rapidly with increased temperature to $T_1 = 5.22 \pm 1.54 \units{\mu s}$ at 5 K. Above 5 K the A1 and B2 transitions broaden and blur together, inhibiting qubit initialization and readout. Below 3 K, the long $T_1$ timescales are slow to measure due to limited readout fidelity. At 1.7 K, no appreciable $T_1$ decay was observed for up to $20 \units{ms}$.

Measurements of the log of $T_1$ versus temperature are fit to the following model \cite{wolfowicz:2021}:
\begin{equation}
    \Gamma_\mathrm{ph}(T) \propto \frac{\Delta_g^3}{e^{\hbar \Delta_g / k_B T}-1}, \\
    \label{eqn:T1_vs_temp_model}
\end{equation}
where $\Gamma_\mathrm{ph}(T)/2\pi$ is based on the rate of phonon-induced transitions predicted for group IV centers in diamond \cite{jahnke:2015,pingault:2017,sohn:2018,meesala:2018},
with $1/T_1 = \Gamma_\mathrm{ph}(T)/2\pi$. In this model, $\Gamma_\mathrm{ph}(T)$ scales as the Bose-Einstein distribution dependent on the ground state splitting $\Delta_g/2\pi = 903 \units{GHz}$ (measured in Fig.~\ref{fig:pl}b). 

We note that our measurements are shorter than previous SnV$^-$ measurements of $T_1$ versus temperature (e.g. at 4 K, Ref.~\cite{trusheim:2020} reports $T_1 \approx 1 \units{ms}$ and we report $T_1=81\pm13 \units{\mu s}$), but far longer than similar experiments using SiV$^-$ centers at the same temperature (e.g. Ref.~\cite{pingault:2017} reports $T_1 \approx 300 \units{ns}$ at 4 K). Potentially, this reduction arises from the increased strain on this emitter, similar to work in the SiV$^-$ center \cite{sohn:2018,meesala:2018}, but requires further investigation.

Next, we measure coherence time as a function of temperature. Coherence time $T_2^\mathrm{echo}$ remains near $170\units{\mu s}$ until $3\units{K}$, at which point it begins to decrease.
We fit the log of $T_2^\mathrm{echo}$ to the model:
\begin{equation}
    \Gamma(T) = \Gamma_0 + \Gamma_\mathrm{ph}(T),
    \label{eqn:T2_vs_temp_model}
\end{equation}
where $\Gamma_0$ is a constant dephasing rate, $\Gamma_\mathrm{ph}(T)$ is given by Eq.~\ref{eqn:T1_vs_temp_model}, and $1/T_2^\mathrm{echo} = \Gamma(T)/2\pi$. Fitting to this model gives $\Gamma_0/2\pi = 5.59 \pm 0.56 \units{kHz}$.

The models of $T_1$ and $T_2^\mathrm{echo}$ follow the data in Fig.~\ref{fig:t2_vs_temp} but with some discrepancy. In Fig.~\ref{fig:t2_vs_temp} we also include phenomenological models (thin solid lines) where a factor $\alpha$ is added to scale the term $\Delta_g/T$, such that such that a modified version of Eq.~\ref{eqn:T1_vs_temp_model} reads $\Gamma_\mathrm{ph}(T) \propto \Delta_g^3/(e^{\alpha\hbar\Delta_g/k_B T}-1)$. We fit $\alpha$ as a free parameter, and find $\alpha=1.208\pm0.045$ when fitting to the measurement of $T_1$ versus temperature, or $\alpha=1.207\pm0.029$ when fitting to $T_2^\mathrm{echo}$ versus temperature. These fits follow our data more closely, and could indicate systematic error in our determination of $\Delta_g$ and/or temperature. For example, $\alpha\approx1.2$ corresponds to a temperature of $\approx1.4 \units{K}$ (instead of $1.7 \units{K})$ or a ground state splitting of $\approx1100 \units{GHz}$ (instead of $903 \units{GHz}$). Alternatively, $\alpha\neq1$ could simply indicate underlying models that differ from Eq.~\ref{eqn:T1_vs_temp_model} and Eq.~\ref{eqn:T2_vs_temp_model}.

Regardless, the study of coherence versus temperature in Fig.~\ref{fig:t2_vs_temp} shows that the SnV$^-$ has remarkable potential as a spin qubit. Extrapolating the fit of energy relaxation to low temperatures yields $T_1 \approx 200 \units{s}$ at 1.7 K. Extrapolating the fit of coherence time to 1.7 K yields a temperature-limited Hahn echo $T_2^\mathrm{echo} \approx 1.3 \units{s}$, assuming the magnetic noise induced dephasing rate $\Gamma_0$ is reduced to zero. Dynamical decoupling should therefore be able to extend coherence into the seconds regime if drive heating can be mitigated. 

\section{Conclusion}
In conclusion, we demonstrate high-fidelity microwave control of the ground state spin of a single SnV$^-$ center in diamond. We achieve control fidelity of $99.51\% \pm 0.03\%$ for a $\pi$ pulse and $95.04\% \pm 0.14\%$ for a random Clifford gate. Furthermore, we quantitatively understand that microwave control results from use of our strained SnV$^-$. The measured rate of microwave control matches the expected rate given the experiment geometry and the ground state strain of $\Upsilon_g/2\pi = 177.7 \pm 1.4 \units{GHz}$, which we independently characterize via spectroscopy of the SnV$^-$ level structure. Our drive wire geometry can also be easily improved, boosting future Rabi rates. Crucially, strain does not break inversion symmetry and the measured SnV$^-$ retains stable, narrow, optical lines.

Using high-fidelity pulses we measure a coherence time of $T_2^\mathrm{echo} = 170.0 \pm 2.8 \units{\mu s}$ using a Hahn-echo sequence. We show that coherence can be extended using more decoupling pulses, for example to $T_2^\mathrm{XY16} = 650.2 \pm 28.4 \units{\mu s}$ using an XY16 sequence. We confirm the dominant role of paramagnetic defects in the Hahn-echo decoherence time of this system, while explaining the observed Ramsey decay time arising from a probable natural variation of the nuclear spin bath from defect-to-defect. We connect the role of the electron-spin bath to the dependence of the coherence times with applied magnetic field, and simulate the effect of instantaneous diffusion for this system. These results imply a trade-off between magnetic field alignment and coherence. Understanding the coherence of this system points toward future improvements by both isotopic engineering of the diamond and reduction of damage-induced electronic spin states.

Finally, we measure qubit coherence as a function of temperature and find that $T_2^\mathrm{echo} > 100 \units{\mu s}$ at temperatures below 3 K. This gives promise that drive-induced heating can be minimized at temperatures accessible with a helium bath cryostat; consistent with our demonstrated ability to apply many high-fidelity gates. Extrapolating the measured temperature dependence suggests that coherence can be on the order of seconds at 1.7 K, so long as other sources of dephasing (nuclear and electronic spin baths and drive-induced heating), can be sufficiently eliminated.

Together, these results characterize the control and coherence of the SnV$^-$ qubit in diamond. In particular, we show that the SnV$^-$ is an attractive spin qubit with high-fidelity gates and long coherence times at 1.7 K. Combined with other recent SnV$^-$ advances including nanophotonic integration \cite{rugar:2021,parker:2023}, single-photon indistinguishability \cite{martinez:2022}, single-shot nuclear spin readout \cite{parker:2023}, and spectroscopy of the hyperfine structure \cite{parker:2023,harris:2023}, the SnV$^-$ is now an increasingly well understood and favorable platform for building the next generation of quantum networks. 

\vspace{0.1in}
\textit{Acknowledgments.}--- This work has been supported by the Department of Energy under the Q-NEXT program, and Grants No. DE-SC0020115 and No. DE-AC02-76SF00515. E.I.R. and C.P.A. acknowledge support by an appointment to the Intelligence Community Postdoctoral Research Fellowship Program at Stanford University administered by Oak Ridge Institute for Science and Education (ORISE) through an interagency agreement between the U.S. Department of Energy and the Office of the Director of National Intelligence (ODNI). H.L. and H.C.K. acknowledge support by the Burt and Deedee McMurtry Stanford Graduate Fellowship (SGF). J.G. acknowledges support from the Hertz Fellowship. G.S. and S.A. acknowledge support from the Stanford Bloch Postdoctoral Fellowship. D.R. acknowledges support from the Swiss National Science Foundation (Project No. P400P2\_194424). K.V.G is supported by the BAEF and the FWO (12ZB520N). D.R. and S.A. contributed to this work prior to joining AWS. 

We thank Jesús Arjona Martínez, Cathryn Michaels, Ryan Parker, Mykyta Onizhuk, Souvik Biswas, Laura Orphal-Kobin, Tim Schr{\"o}der, Gerg\H{o} Thiering, P\'{e}ter Udvarhelyi, \'{A}d\'{a}m Gali and Joonhee Choi for helpful discussions. We thank Tom Lee for lending a microwave power meter. We thank Daniil Lukin and Alex White for help with printed circuit board preparation. We thank Jacob Feder, Jonathan Marcks, and Mia Froehling Gallier for help with instrument control code, based on the ``nspyre'' framework \cite{feder:2023}. We thank Haiyu Lu, Shuo Li, Patrick McQuade, Zhi-Xun Shen and Nicholas Melosh for assistance with diamond sample preparation. We thank Nazar Delgan, F. Joseph Heremans, Michael Titze, and Edward Bielejec for collaboration on the preparation of related devices.

\vspace{0.1in}
\textit{Note added.}—After submission, a related manuscript was independently posted~\cite{guo:2023}.



\twocolumngrid 



\appendix
\label{appendix}
\section{Coherent spin control above 1 K: a motivation for SnV$^-$'s}
\label{sec:SnV_motivation}

Group IV centers in diamond have a ground state manifold whose two lowest states $\ket{1}$ and $\ket{2}$ are operated as a qubit. Because of spin-orbit coupling, these states are separated from the next lowest, $\ket{3}$ and $\ket{4}$, by the ground state splitting $\Delta_g/2\pi$. Long qubit coherence requires operation at a low enough temperature to avoid phonon-mediated transitions between these states. These transitions excite the system to the $\ket{3}$ and $\ket{4}$ levels at the temperature dependent rate $\Gamma_\mathrm{ph}(T)$.

\begin{figure}[htb!] 
\begin{center}
\includegraphics[width=1.0\columnwidth]{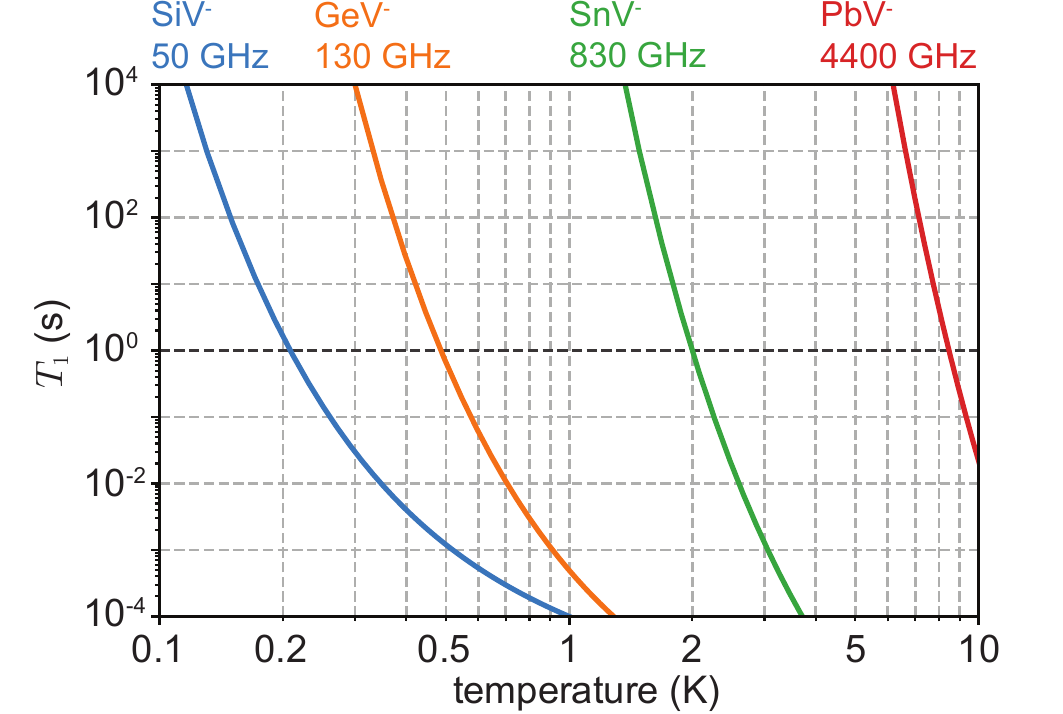}
\caption{
Model of qubit energy relaxation time, $T_1$, as a function of temperature for the group IV centers in diamond. Model is based on the rate of phonon-induced transitions Ref.~\cite{jahnke:2015} and Eq.~\ref{eqn:T1_vs_temp_model}. Here we assume $\Gamma_\mathrm{ph}/2\pi = 1/T_1$ and use the proportionality constant obtained from a fit to our measurement in Fig.~\ref{fig:t2_vs_temp}. Darker horizontal line is 1 s.
}
\label{fig:group4_phonon_dephasing}
\end{center}
\end{figure}

The rate $\Gamma_\mathrm{ph}(T)$ is predicted to scale with temperature according to the cube of the ground state splitting times the Boltzmann distribution, Eq.~\ref{eqn:T1_vs_temp_model} \cite{jahnke:2015}. This limits the qubit's energy relaxation time, $T_1(T) = 2\pi/\Gamma_\mathrm{ph}(T)$. In Fig.~\ref{fig:group4_phonon_dephasing} we model $T_1(T)$ versus temperature for the group IV centers in diamond.

We plot this model in Fig.~\ref{fig:group4_phonon_dephasing} for all of the group IV centers in diamond: silicon (SiV$^-$), germanium (GeV$^-$), tin (SnV$^-$) and lead (PbV$^-$). These simulations all assume the proportionality constant obtained from the fit to our measurement of $T_1$ versus temperature in Fig.~\ref{fig:t2_vs_temp}.

All group IV centers are predicted to have long $T_1$ times the low temperature limit, but the temperature threshold at which long lifetime occurs increases for greater ground state splitting. From this model we predict a $T_1$ time of 1 s for the SiV$^-$ at 0.2 K, the GeV$^-$ at 0.5 K, the SnV$^-$ at 2 K, and the PbV$^-$ at 8 K.

The SnV$^-$'s comparatively large ground state splitting is advantageous because greater cooling power is available at elevated temperatures. This is especially important for microwave spin control, whose main limitation is drive-induced heating. Commercially available dilution refrigerators today (e.g. Oxford and Bluefors) provide about 0.5 mW of cooling power around 100 mK, around where spin coherent SiV$^-$'s must operate. 

In contrast, hundreds of milliwatts of cooling power are available above 1 K \cite{wang:2014}. For instance, evaporative bath cryostats may be used above 0.3 K (circulating $^3\mathrm{He}$) or above 1.3 K (circulating $^4\mathrm{He}$). The cooling power $\dot{Q}^\mathrm{evap}$ of such cryostats scales exponentially with temperature as \cite{pobell:2007} $\dot{Q}^\mathrm{evap} \propto e^{-L/R T}$, where, $R = 8.3145 \units{J \cdot mol^{-1} \cdot K^{-1}}$ is the ideal gas constant and $L$ is the gas' latent heat (between $85-91 \units{J/mol}$ for $^4\mathrm{He}$). This scaling is faster than that of the helium dilution process, which scales quadratically in the low temperature limit and only works up to 0.87 K. 

In conclusion, there is vastly more cooling power available for SnV$^-$ based experiments which can operate above 1 K, compared to similar SiV$^-$ experiments which must operate at millikelvin temperatures. This will reduce the burden of drive-induced heating, leading to higher control fidelities and more scalable experiments.

\section{SnV$^-$ Model}
\label{sec:model}
In this appendix we model the SnV$^-$ center in diamond in order to explain a central feature of our work: that strain is advantageous for spin control. We begin with a summary of the SnV$^-$ Hamiltonian, drawing upon the work in Refs.~\cite{hepp:2014,thiering:2018,nguyen:2019,trusheim:2020,debroux:2021} and others. We then numerically simulate the SnV$^-$'s experimentally relevant properties: eigenstates, transition frequencies, branching ratio, microwave Rabi rate, etc., as functions of strain and applied magnetic field.

\subsection{Undriven Hamiltonian}
\label{sec:undriven_hamiltonian}
The SnV$^-$ is a spin-1/2 center with degrees of freedom $\ket{\uparrow}$ (spin up) and $\ket{\downarrow}$ (spin down). The SnV$^-$ also has orbital degree of freedom $\ket{e_x}$ and $\ket{e_y}$, here expressed in the ``\textit{x/y}'' basis where $x$ and $y$ are coordinates relating to the spatial orientation of the center's orbitals with respect to the lattice \cite{hepp:2014,trusheim:2020,debroux:2021}. These orbital and spin degrees of freedom combine so that the SnV$^-$ has eigenstates $\{ \ket{e_x \uparrow}, \ket{e_x \downarrow}, \ket{e_y \uparrow}, \ket{e_y \downarrow}  \}$.

The Hamiltonian of an SnV$^-$ is composed of two orthogonal subspaces describing the ground and excited state manifolds. These have Hamiltonians $\hat{H}_\mathrm{SnV^-}^g$ (ground) and $\hat{H}_\mathrm{SnV^-}^e$ (excited), denoted by superscripts $g$ and $e$, respectively. At zero magnetic field the Hamiltonian of each subspace is dominated by spin-orbit coupling $\hat{H}_\mathrm{SO}^{g,e}$ and the Jahn-Teller effect (indistinguishable from strain) $\hat{H}_\mathrm{JT}^{g,e}$ \cite{hepp:2014,thiering:2018}:

\begin{equation}
    \hat{H}_\mathrm{SO}^{g,e} = 
    -\frac{\hbar \lambda_{g,e}}{2}
    \begin{bmatrix}
    0 & i \\
    -i & 0 \\
    \end{bmatrix}
    \otimes
    \begin{bmatrix}
    1 & 0 \\
    0 & -1 \\
    \end{bmatrix},  
    \label{eqn:HSO}     
    \tag{B1}
\end{equation}

\begin{equation}
    \hat{H}_\mathrm{JT}^{g,e} = \hbar
    \begin{bmatrix}
    \Upsilon^x_{g,e} & \Upsilon^y_{g,e} \\
    \Upsilon^y_{g,e} & -\Upsilon^x_{g,e} \\
    \end{bmatrix}
    \otimes
    \begin{bmatrix}
    1 & 0 \\
    0 & 1 \\
    \end{bmatrix}.
    \label{eqn:HJT}    
    \tag{B2}
\end{equation}
Here $\lambda_{g,e}$ is the strength of spin-orbit coupling, and $\Upsilon_{g,e}^x$ and $\Upsilon_{g,e}^y$ are the transverse components of Jahn-Teller or strain effects, where the axial component of strain (an identity term in the orbital component) is neglected because it leads to common mode energy shifts only. Note that the magnitude of both spin-orbit coupling and strain may differ between the ground and excited state manifolds.

The Zeeman effect modifies the SnV$^-$ Hamiltonian under an applied magnetic field by: 
\begin{multline}    
    \hat{H}_\mathrm{Z}^{g,e} = \\
    \frac{\hbar \gamma}{2}
    \begin{bmatrix}
    1 & 0 \\
    0 & 1 \\
    \end{bmatrix}
    \otimes
    \begin{bmatrix}
    (1+2\delta_{g,e}) B_{\parallel} & B_{\perp} \\
    B_{\perp}^{*} & -(1+2\delta_{g,e}) B_{\parallel} \\
    \end{bmatrix},
    \label{eqn:HB}    
    \tag{B3}
\end{multline}
where $\gamma/2\pi \approx 28 \units{GHz/T}$ is the electron gyromagnetic ratio.
In Eq.~\ref{eqn:HB}, $B_{\parallel} = B_z$ and $B_{\perp} = B_x + i B_y$ are components of the external static magnetic field $\Vec{B} = \{B_x, B_y, B_z\}$, in the coordinate frame oriented along the spin's dipole $\Vec{\mu}$. The factor $\delta_{g,e}$ describes anisotropy of the spin's $g$ factor, and is predicted from \textit{ab initio} calculations \cite{thiering:2018}. The Zeeman effect also has the following orbital contribution \cite{hepp:2014,trusheim:2020}:
\begin{equation}
    \hat{H}_\mathrm{L}^{g,e} = 
    \frac{\hbar \gamma f_{g,e}}{2}
    \begin{bmatrix}
    0 & i B_{\parallel} \\
    -i B_{\parallel} & 0 \\
    \end{bmatrix}
    \otimes
    \begin{bmatrix}
    1 & 0 \\
    0 & 1 \\
    \end{bmatrix},
    \label{eqn:HL}    
    \tag{B4}
\end{equation}
where $f_{e,g}$ is the quenching factor.

We therefore model the total SnV$^-$ Hamiltonian as
\begin{equation}
\hat{H}_\mathrm{SnV^-}^{g,e} = \hat{H}_\mathrm{SO}^{g,e} + \hat{H}_\mathrm{JT}^{g,e} + 
\hat{H}_\mathrm{Z}^{g,e} +
\hat{H}_\mathrm{L}^{g,e}.
\label{eqn:HSnV}
\tag{B5}
\end{equation}
As graphically illustrated in Fig.~\ref{fig:intro}b, we denote the eigenstates of $\hat{H}_\mathrm{SnV^-}^{g}$ as $\{ \ket{1}, \ket{2}, \ket{3}, \ket{4}\}$ and the eigenstates of $\hat{H}_\mathrm{SnV^-}^{e}$ as $\{ \ket{A}, \ket{B}, \ket{C}, \ket{D} \}$. 

At zero magnetic field the eigenstates of $\hat{H}_\mathrm{SnV^-}^{g,e}$ have splitting $\Delta_g$, the difference in angular frequency between the degenerate states $\ket{1}$ and $\ket{2}$ versus $\ket{3}$ and $\ket{4}$, and an excited state splitting $\Delta_e$, the difference in angular frequency between the degenerate states $\ket{A}$ and $\ket{B}$ versus $\ket{C}$ and $\ket{D}$. Splitting depends on spin-orbit coupling and strain as follows: 
\begin{equation}
    \Delta_{g,e} = \sqrt{\lambda_{g,e}^2 + 4\Upsilon_{g,e}^2},
    \label{eqn:ground_state_splitting}
    \tag{B6}
\end{equation}
where $\Upsilon_{g,e} = \sqrt{(\Upsilon_{g,e}^x)^2 + (\Upsilon_{g,e}^y)^2}$ is the magnitude of strain, which throughout this work we refer to as just strain. (Note that in other literature, e.g. \cite{meesala:2018}, ``strain'' may instead refer to dimensionless tensor which is related to $\Upsilon_{g,e}^x$ and $\Upsilon_{g,e}^y$ via a coupling constant.)

\subsection{Qubit frequency under a perpendicular field}
The qubit angular frequency is $\omega_q = \omega_2 - \omega_1$, where $\omega_1/2\pi$ and $\omega_2/2\pi$ are the eigenfrequencies of the $\ket{1}$ and $\ket{2}$ states, respectively. Here we consider $\omega_q$ in the limit of nonzero strain and where a nonzero magnetic field is applied perpendicular to the spin axis only (i.e., $B_\perp\neq0$ and $B_\parallel=0$). Under these conditions,
\begin{align}
    \omega_1 = -\frac{1}{2} \sqrt{\lambda_g^2 + (\gamma B_\perp + 2\Upsilon_g)^2}, \tag{B7} \\
    \omega_2 = -\frac{1}{2} \sqrt{\lambda_g^2 + (\gamma B_\perp - 2\Upsilon_g)^2}. \tag{B8}
\end{align}

The expression for qubit frequency can now be simplified. Assuming $\gamma B_\perp \ll \Upsilon_g, \lambda_g$ and Taylor expanding in terms of $\gamma B_\perp$ gives:
\begin{equation}
    \omega_q = \frac{2 \gamma B_\perp \Upsilon_g}{\Delta_g} + \mathcal{O}(\gamma B_\perp)^3 + \cdots.
    \tag{B9}
    \label{eqn:f_qubit}
\end{equation}
Therefore, to first order the qubit frequency scales as $\Upsilon_g / \Delta_g$ or alternatively, $\Upsilon_g \approx \omega_q \Delta_g / 2 \gamma B_\perp$. Equation~\ref{eqn:f_qubit} thus gives a convenient way to measure strain using measurements of $\Delta_g$ (Fig.~\ref{fig:pl}b) and $\omega_q$ (Fig.~\ref{fig:ple}a).

\subsection{Driven Hamiltonian}

An SnV$^-$'s state can be controlled by applied electromagnetic drives with Hamiltonian $\hat{H}^\prime$. This turns on an interaction between two eigenstates $\ket{\psi_1}$ and $\ket{\psi_0}$, with the transition matrix element $\bra{\psi_1} \hat{H}^\prime \ket{\psi_0}$.

\textit{Microwave driving}.--- An oscillating magnetic field at the frequency difference between two eigenstates creates the interaction Hamiltonian: 
\begin{equation}
    \hat{H}_\mathrm{MW} = \frac{\hbar \gamma}{2}
    \begin{bmatrix}
    1 & 0 \\
    0 & 1 \\
    \end{bmatrix}
    \otimes
    \begin{bmatrix}
    b_{\parallel} & b_{\perp} \\
    b_{\perp}^* & -b_{\parallel} \\
    \end{bmatrix},
    \label{eqn:HMW}
    \tag{B10}
\end{equation}
where $\vec{b} = \{b_x, b_y, b_z\}$ is the microwave magnetic field such that $b_{\parallel} = b_z$ and $b_\perp = b_x + i b_y$. Note that Eq.~\ref{eqn:HMW} neglects any effect of the microwave drive on the orbital degree of freedom due to the orbital Zeeman effect and neglects any anisotropy of the spin's dipole moment. In other words, Eq.~\ref{eqn:HMW} makes the assumption that $f_g = 0$ and $\delta_g = 0$. Notice also that Eq.~\ref{eqn:HMW} has the same form as Eq.~\ref{eqn:HB}; summed together they describe the emitter's interaction with any external magnetic field. For clarity we simply separate the effect of a static field $\vec{B}$, and an oscillating field $\vec{b}$. 

The Rabi rate $\Omega_\mathrm{MW}$ between states $\ket{1}$ and $\ket{2}$ under an applied microwave drive is the transition dipole element:
\begin{equation}
    \Omega_\mathrm{MW} = \frac{1}{\hbar} \bra{2} \hat{H}_\mathrm{MW} \ket{1}.
    \label{eqn:rabi_mw}
    \tag{B11}
\end{equation}

\textit{Optical driving}.--- Optical light acts on the orbital degree-of-freedom with the interaction Hamiltonian $\hat{H}' = -\hat{p} \cdot \vec{E}$, where $\hat{p} = \{ p_x, p_y, p_z \}$ is the dipole operator and $\vec{E} = \{ E_x, E_y, E_z \}$ is the electric field of the incident or emitted light. The dipole operator is defined as $\hat{p} = e \hat{r}$, where $e$ is the charge of an electron and $\hat{r}$ is the position operator. The position operator is well approximated by a $\delta$ function at the emitter's location, since its spatial extent is much smaller than the diffraction limited size of the laser excitation.

For unpolarized light, the dipole operator for the transitions between the excited state level $\ket{A}$ and ground state levels $\ket{1}$ and $\ket{2}$ has the following orbital components, defined in the ``\textit{x/y}'' basis as \cite{hepp:2014}:
\begin{equation}
    \hat{p}_x =
    q \begin{bmatrix}
    1 & 0 \\
    0 & -1 \\
    \end{bmatrix},
    \quad
    \hat{p}_y =
    q \begin{bmatrix}
    0 & -1 \\
    -1 & 0 \\
    \end{bmatrix},
    \quad
    \hat{p}_z =
    q \begin{bmatrix}
    1 & 0 \\
    0 & 1 \\
    \end{bmatrix},
    \label{eqn:dipole_operators}    
    \tag{B12}
\end{equation}
which act as the identity in the spin basis, and where $q$ is the charge of an electron. Note that we do not express the dipole operators as the tensor product of two $2\times2$ matrices, as with other Hamiltonians in this appendix, because they describe an interaction between the ground and excited state manifolds.

From Eq.~\ref{eqn:dipole_operators} we compute $P_\mathrm{A1}$ and $P_\mathrm{A2}$, the probability at which population transfers along the spin conserving A1 transition (between $\ket{1}$ and $\ket{A}$) or the spin flipping A2 transition (between $\ket{2}$ and $\ket{A}$), respectively. Under a resonant drive these probabilities are proportional to \cite{hepp:2014}
\begin{equation}
    P_\mathrm{A1} \propto \left| \bra{A} \hat{p} \cdot \Vec{E} \ket{1} \right|^2,
    \label{eqn:Gamma_A1}
    \tag{B13}
\end{equation}
\begin{equation}
    P_\mathrm{A2} \propto \left| \bra{A} \hat{p} \cdot \Vec{E} \ket{2} \right|^2,
    \label{eqn:Gamma_A2}
    \tag{B14}
\end{equation}
with $\hat{p} \cdot \Vec{E} = \hat{p}_x E_x + \hat{p}_y E_y + \hat{p}_z E_z$. 

The ratio between these rates is the branching ratio $\eta$:
\begin{equation}
    \eta = P_\mathrm{A1} / P_\mathrm{A2}.
    \label{eqn:branching_ratio}
    \tag{B15}
\end{equation}
This ratio is the likelihood that, when excited into state $\ket{A}$, the emitter will decay via emission of a photon at the spin preserving transition (A1) versus emission of a photon at the spin flipping transition (A2). The emitter may alternatively emit a lower energy (higher wavelength) photon into its phonon sideband.

The branching ratio, also known as the ``cyclicity'', is relevant to qubit state initialization and readout. Under a drive resonant with a spin preserving transition (e.g., A1), a higher branching ratio (greater cyclicity) means slower qubit state initialization. But it also causes more photons to be emitted before the spin state is destroyed, thus improving qubit readout. A lower branching ratio (lower cyclicity) means faster spin state initialization, but reduced photons emitted and thus reduced readout signal.

\begin{figure}[tb!] 
\begin{center}
\includegraphics[width=1.0\columnwidth]{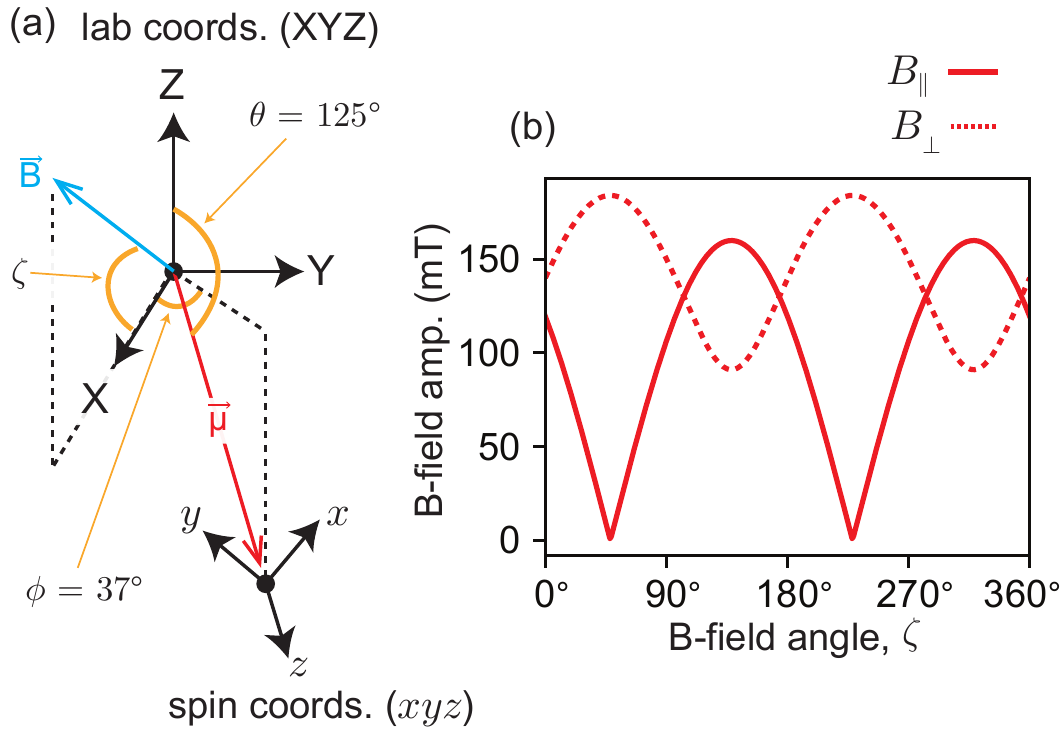}
\caption{
(a) Lab coordinates: axes X, Y, Z are coordinates in the lab frame, with X and Z set by the orientation of the magnet. The magnet applies field $\Vec{B}$ with components $B_\mathrm{X} = |\Vec{B}|\cos{(\zeta)}$ and $B_\mathrm{Z} = |\Vec{B}|\sin{(\zeta)}$. A vector in the lab frame is defined by polar angle $\theta$ and azimuthal angle $\phi$. Spin coordinates: axes $x$, $y$ and $z$ are defined so that $z$ is parallel to the dipole moment of the spin, which is oriented along vector $\vec{\mu}$ in the lab frame.
(b) Illustration of the B-field angular sweeps in Figs.~\ref{fig:ple}a~and~\ref{fig:sims_vs_strain}. Magnetic field is swept along angle $\zeta$, changing its orientation on the circle formed by the X/Z axes. $B_\parallel$ and $B_\perp$ are the field component parallel to and perpendicular to, respectively, the spin's dipole moment $\vec{\mu}$.
}
\label{fig:coordinates}
\end{center}
\end{figure}

\subsection{Hamiltonian determination}
\label{sec:H_determination}

In this section we determine parameters of the SnV$^-$ Hamiltonian, Eq.~\ref{eqn:HSnV}, using a fit to the spectroscopic measurements in Fig.~\ref{fig:ple}a. In particular, this lets us fit the anisotropy $\delta_{g,e}$ of the SnV$^-$ $g$ factor and the strength of the orbital Zeeman term $f_{g,e}$. These parameters have not yet been precisely measured for the SnV$^-$ center. Fitting our spectroscopic measurements also allows us to extract the exact magnitude of strain and spin-orbit terms.

The parameters $\delta_{g,e}$ and $f_{g,e}$ are all related to the orbital reduction factors $g_{L}^g$ (ground state) and $g_{L}^e$ (excited state) by \cite{thiering:2018}:
\begin{align}
    f_{g,e} = g_L^{g,e} p^{g,e}, \label{eqn:f} \tag{B16} \\
    \delta_{g,e} = g_L^{g,e} \delta_p^{g,e}. \tag{B17} \label{eqn:delta}
\end{align}
Here, $\delta_p^{g,e}$ and $p^{g,e}$ are reduction parameters, which are obtained by the \textit{ab initio} calculations in Ref.\cite{thiering:2018} to have values $\delta_p^g=0.042$ , $\delta_p^e=0.303$, $p^{g}=0.471$, and $p^{e}=0.125$. On the other hand, in Ref.~\cite{thiering:2018} the quoted values of the orbital reduction factors $g_{L}^{g,e}$ are estimated from measurements of the SiV$^-$ center \cite{hepp:2014,thiering:2018}. This uncertainty about $g_{L}^{g,e}$ for the SnV$^-$ leads to uncertainty in both $f_{g,e}$ and $\delta_{g,e}$ beyond the uncertainty from \textit{ab initio} calculations used to compute $\delta_p^{g,e}$ and $p^{g,e}$.

To overcome this uncertainty, we use the orbital reduction factors $g_L^{g}$ and $g_L^{e}$ as free parameters in the fit of the measurements in Fig.~\ref{fig:ple}a, while using the computed $\delta_p^{g,e}$ and $p^{g,e}$. We then obtain measurements of $f_g$, $f_e$, $\delta_g$, and $\delta_e$ using Eqs.~\ref{eqn:f}~and~\ref{eqn:delta}, respectively. These results are summarized in Table ~\ref{tab:SnVparameters}. We note that the computed $\delta_p^{g,e}$ and $p^{g,e}$ should be fairly accurate because the same calculations in Ref.~\cite{thiering:2018} accurately compute the spin-orbit strength.

\begin{table}[tb!]
\caption{
SnV$^-$ parameters predicted from \textit{ab initio} calculations are compared to measured values. The ground state splitting $\Delta_g/2\pi$ is measured directly from a photoluminescence (PL) measurement, Fig.~\ref{fig:pl}b. The following terms are fit based on the spectroscopic measurements in Fig.~\ref{fig:ple}: $\delta_g$ and $\delta_e$ (anisotropic Zeeman), $\phi$ (azimuthal angle of spin dipole moment), $\Upsilon_g/2\pi$ and $\Upsilon_e/2\pi$ (strain magnitude) and $f_g$ and $f_e$ (orbital Zeeman). Note that the results obtained from $\delta_g$, $\delta_e$, $f_g$ and $f_e$ are determined via a fit to the orbital reduction factors $g_{L}^g = 0.363 \pm 0.009$ and $g_{L}^e = 0.581 \pm 0.009$, multiplied by the parameters $\delta_p^{g,e}$ and $p^{g,e}$ which are obtained from \textit{ab initio} calculations (see Eqs.~\ref{eqn:f}~and~\ref{eqn:delta}). The excited state splitting $\Delta_e/2\pi$ is fixed using values expected from the literature \cite{thiering:2018}.
Spin-orbit coupling ($\lambda_g/2\pi$ and $\lambda_e/2\pi$) is determined from ground/excited state splitting and strain using Eq.~\ref{eqn:ground_state_splitting}. Finally $\theta$ (polar angle of spin dipole moment) is fixed, using a value which assumes the Z-axis in lab coordinates is normal to the surface of the chip, to which the spin is oriented at a polar angle expected for an SnV$^-$ in $\langle100\rangle$ diamond \cite{debroux:2021}.
}
  \begin{center}
    \begin{tabular}{ |p{1.0cm}|p{1.75cm}|p{2.75cm}|p{2.5cm}|  }
         \hline
         Term & Ref.~\cite{thiering:2018} & This work & Method \\
         \hline
         \hline
         $\Delta_g/2\pi$ & 850 GHz & $902.98 \pm 0.73 \units{GHz}$ & Fit, Fig.~\ref{fig:pl}b \\
         $\Delta_e/2\pi$ & 3000 GHz & 3000 GHz & Fixed \\
         $\Upsilon_g/2\pi$ &   & $177.67 \pm 1.37 \units{GHz}$ & Fit, Fig.~\ref{fig:ple} data \\
         $\Upsilon_e/2\pi$ &   & $134.00 \pm 12.61 \units{GHz}$ & Fit, Fig.~\ref{fig:ple} data \\
         $\lambda_g/2\pi$ & 850 GHz & $830.15 \pm 1.42 \units{GHz}$ & Eq.~\ref{eqn:ground_state_splitting} \\
         $\lambda_e/2\pi$ & 3000 GHz & $2988.0 \pm 2.26 \units{GHz}$ & Eq.~\ref{eqn:ground_state_splitting} \\
         $g_{L}^g$ & 0.328 & $0.363 \pm 0.009$ & Fit, Fig.~\ref{fig:ple} data \\
         $g_{L}^e$ & 0.782 & $0.581 \pm 0.009$ & Fit, Fig.~\ref{fig:ple} data \\
         $f_g$ & 0.154 & $0.171 \pm 0.004$ & Eq.~\ref{eqn:f} \\
         $f_e$ & 0.098 & $0.073 \pm 0.001$ & Eq.~\ref{eqn:f} \\
         $\delta_g$ & 0.014 & $0.0152 \pm 0.0004$ & Eq.~\ref{eqn:delta} \\
         $\delta_e$ & 0.238 & $0.1761 \pm 0.0028$ & Eq.~\ref{eqn:delta} \\
         $\theta$ &   & $125.3 \units{deg}$ & Fixed \\
         $\phi$ &   & $37.33 \pm 0.47 \units{deg}$ & Fit, Fig.~\ref{fig:ple} data \\
         \hline
    \end{tabular}
  \label{tab:SnVparameters}
  \end{center}
\end{table}

\subsection{Simulated behavior}
\label{sec:simulated_behavior}

In this section, we model how strain and magnetic field affect experimentally relevant properties including the energy spectrum, branching ratio $\eta$, and microwave Rabi rate $\Omega_\mathrm{MW}$. Properties are simulated numerically using Eq.~\ref{eqn:HSnV}, using the parameters listed in Table~\ref{tab:SnVparameters}. 

\begin{figure*}[tb!] 
\begin{center}
\includegraphics[width=2.0\columnwidth]{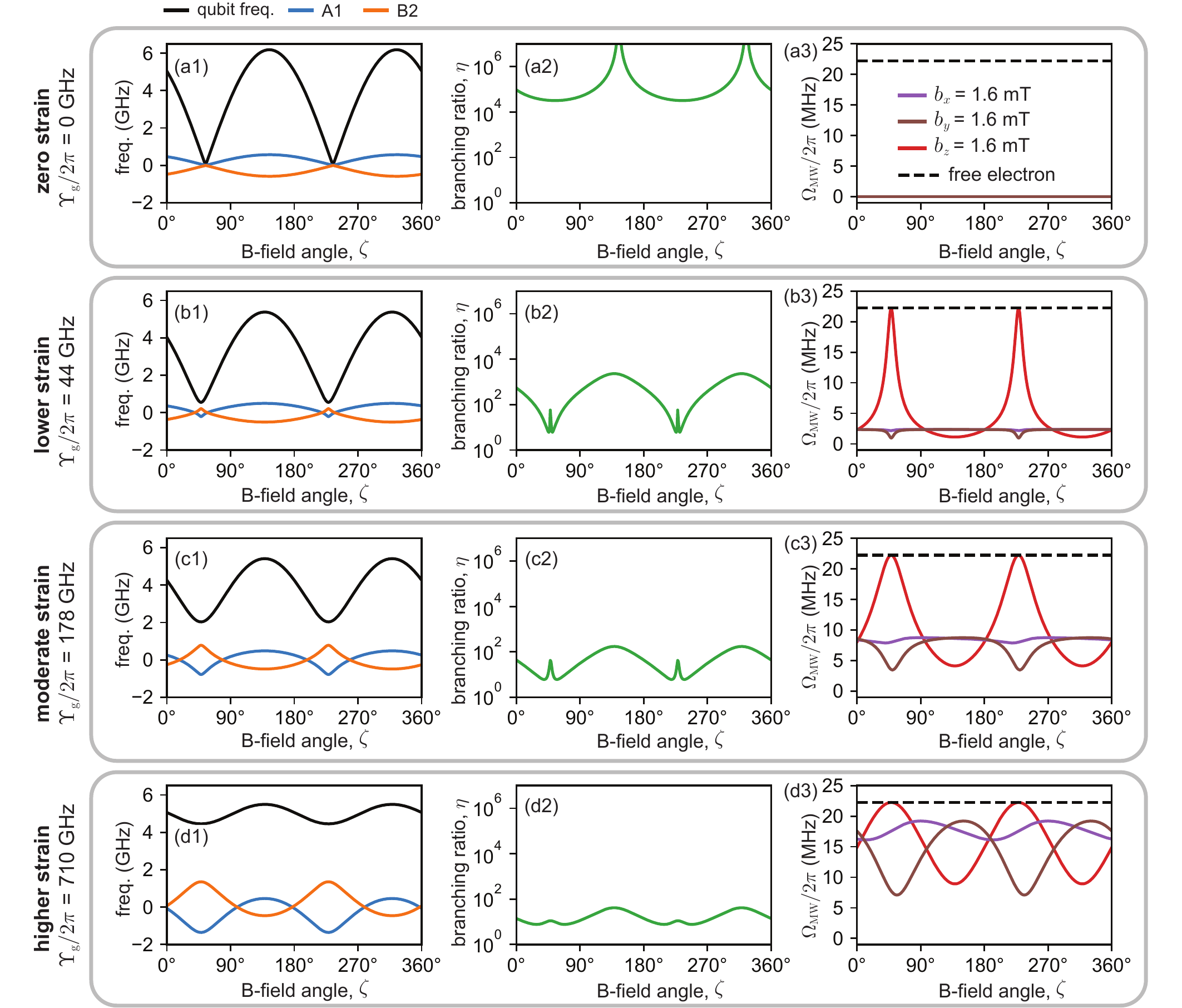}
\caption{
Simulated SnV$^-$ properties versus magnetic field orientation for $|\Vec{B}| = 150 \units{mT}$. B-field angle $\zeta$ is swept along the X/Z axes in the lab frame, Fig.~\ref{fig:coordinates}a, such that $B_\mathrm{X}=|\Vec{B}|\cos(\zeta)$ and $B_\mathrm{Z}=|\Vec{B}|\sin(\zeta)$.
Simulations use the following strain values: (a1)-(a3) zero strain, $\Upsilon_g/2\pi = 0 \units{GHz}$; (b1)-(b3) lower strain, $\Upsilon_g/2\pi = 44.5 \units{GHz}$; (c1)-(c3) moderate strain, $\Upsilon_g/2\pi = 177.7 \units{GHz}$; (d1)-(d3) higher strain, $\Upsilon_g/2\pi = 710.8 \units{GHz}$. The ratio $\Upsilon_g/\Upsilon_e = 1.33$ is assumed constant for all simulations based on the values in this experiment, Table~\ref{tab:SnVparameters}, but may in general change among different SnV$^-$'s. (a1)(b1)(c1)(d1) SnV$^-$ qubit frequency and optical transitions $\mathrm{A1}$ and $\mathrm{B2}$. Optical transitions are plotted detuned from their mean frequency near 484 THz. (a2)(b2)(c2)(d2) Branching ratio $\eta$, Eq.~\ref{eqn:branching_ratio}. (a3)(b3)(c3)(d3) Microwave Rabi rate $\Omega_\mathrm{MW}/2\pi$, Eq.~\ref{eqn:rabi_mw}. Results are plotted for a drive field either perpendicular to the spin dipole moment (purple, $b_x$ only, or brown, $b_y$ only, with $b_\perp = b_x + i b_y$) or parallel to the SnV$^-$'s dipole moment (red, $b_\parallel = b_z$ only), which is oriented along vector $\Vec{\mu}$ in Fig.~\ref{fig:coordinates}a. Black dashed line is the Rabi rate expected for a free electron driven by an optimally oriented microwave field of the same amplitude. 
}
\label{fig:sims_vs_strain}
\end{center}
\end{figure*}

Results are shown in Fig.~\ref{fig:sims_vs_strain} for four different strain limits: ``zero'' (unstrained), ``lower,'' ``moderate,'' and ``higher''. The ``moderate'' case has strain values chosen to match this experiment. The lower (higher) simulations have $\Upsilon_{g,e}$ decreased (increased) by a factor of 4 in comparison.

SnV$^-$ properties are dependent on both strain and the angular orientation $\zeta$ of $\Vec{B}$ compared to the spin dipole moment $\Vec{\mu}$. At low strain, Fig.~\ref{fig:sims_vs_strain}a, qubit frequency changes dramatically with $\zeta$, and approaches zero as $\Vec{B}\cdot\Vec{\mu}=0$. At higher strain, the qubit frequency becomes less sensitive to $\zeta$ and its minimum value increases.

Cyclicity of the SnV$^-$'s optical transitions $\eta$ is also dependent on strain. For a given excitation power, lower $\eta$ leads to faster spin state initialization but less photoluminescence, and thus reduced spin readout contrast. We calculate that $\eta$ is greatest when strain is lowest, and that for a given strain $\eta$ is greatest when $\Vec{B}\cdot\Vec{\mu}$ is maximized (parallel orientation). We also find that $\eta \approx 82$ at the angle of $\zeta = 110^{\degree}$ used in the main text data, for the moderate strain values of the device used in this experiment, Fig.~\ref{fig:sims_vs_strain}c2. We note that the cyclicity has a small and narrow increase when the field is maximally perpendicular to the spin-orbit axis. In this case, the new quantization for the spin is along the axis on which strain acts, such that spin becomes more conserved and slightly cycling in this new basis. This is evidenced by this feature being absent in the low strain case, and becoming dominant when strain becomes large in comparison to the spin-orbit interaction.

The microwave Rabi rate $\Omega_\mathrm{MW}$ is dependent on strain, B-field angle $\zeta$, and the microwave drive field, $\Vec{b}$. At low strain, $\Omega_\mathrm{MW}$ is maximized when $\Vec{b}$ is oriented parallel to the spin's dipole moment. At higher strain, $\Omega_\mathrm{MW}$ increases but maintains some dependence on drive angle versus external field angle. At zero strain $\Omega_\mathrm{MW} = 0$ for any drive configuration. We note that due to the orbital mixing, that ac magnetic fields both parallel and orthogonal to the spin-orbit axis can drive qubit transitions, depending on the strain. For an emitter with greater strain, the microwave Rabi rate comes closer to that expected for a free electron, Fig.~\ref{fig:sims_vs_strain}. Interestingly, even for very small values of strain, strong microwave driving is possible when the magnetic field is perpendicular to the spin-orbit axis. Unfortunately, in this case, the cyclicity is heavily suppressed and the optical transitions are not resolved. As a result, a trade-off exists between the possible Rabi rate and cyclicity of the transitions.

Using parameters chosen to match the data in Fig.~\ref{fig:rabi} ($\zeta = 110^{\degree}$ and $|\Vec{b}| = 1.6 \units{mT}$), the simulations in Fig.~\ref{fig:sims_vs_strain}c2 predict a Rabi rate between $\Omega_\mathrm{MW}/2\pi \approx 9 \units{MHz}$ for a drive orientation perpendicular to $\Vec{\mu}$, and a Rabi rate of $\Omega_\mathrm{MW}/2\pi \approx 6 \units{MHz}$ for a drive orientation parallel to $\Vec{\mu}$. The amplitude $|\Vec{b}| = 1.6 \units{mT}$ is computed from Ampere's law such that $|\Vec{b}| = \mu_0 I / (2\pi r)$, with $r=63 \units{\mu m}$ the distance from wire bond center to the qubit, $\mu_0$ the permeability of free space, and $I=0.5 \units{A}$ is the microwave drive current corresponding to a continuous microwave power of $\approx 41 \units{dBm}$ passing through the wire bond and $\approx 48 \units{dBm}$ into the cryostat. At these operating conditions we measure a Rabi rate of 20.7 MHz, Fig.~\ref{fig:rabi}, which is consistent with our model to within uncertainty around our determination of the microwave current and distance to the wire bond.

In summary, we simulate the level structure, branching ratio, and microwave Rabi rate of the SnV$^-$ qubit. These simulations use the parameters in Table~\ref{tab:SnVparameters}, which are fit from our measurements in Fig.~\ref{fig:ple}. Simulation results are shown in Fig.~\ref{fig:sims_vs_strain}. We emphasize the following.
\begin{itemize}
    \item Strain is necessary for coherent spin control. More strain increases $\Omega_\mathrm{MW}/2\pi$, all else equal.
    \item More strain decreases cyclicity, harming readout. 
    \item The orientations of static field $\Vec{B}$ and drive field $\Vec{b}$ both affect $\Omega_\mathrm{MW}/2\pi$, but only by a factor of a few at moderate or greater strain.
\end{itemize}

\subsection{Comparison between the SiV$^-$ and SnV$^-$}

Finally, we compare performance of the tin-vacancy center in diamond (SnV$^-$) to the more developed platform of the silicon-vacancy center in diamond (SiV$^-$). These group IV vacancy centers in diamond have similar spin and optical properties but different spin-orbit coupling terms: $\lambda_g/2\pi \approx 50 \units{GHz}$ for the SiV$^-$ and $\lambda_g/2\pi \approx 830 \units{GHz}$ for the SnV$^-$. The ground state splitting depends on both spin-orbit coupling and strain by Eq.~\ref{eqn:ground_state_splitting}.

\begin{figure}[tb!] 
\begin{center}
\includegraphics[width=1.0\columnwidth]{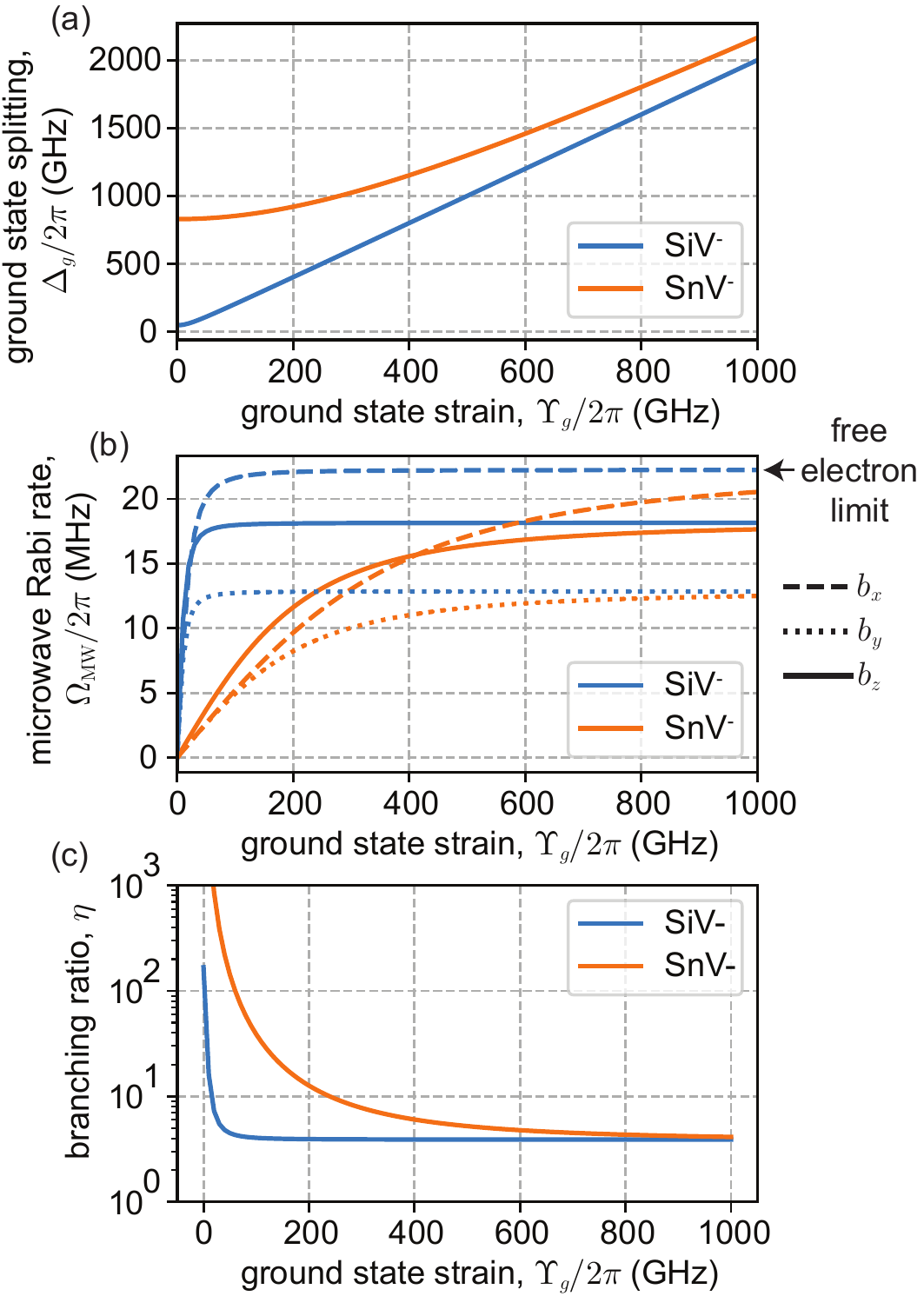}
\caption{
Comparison between the SiV$^-$ and SnV$^-$. (a) Ground state splitting as a function of ground state strain magnitude, Eq.~\ref{eqn:ground_state_splitting}. Simulations assume the SiV$^-$ has a spin-orbit coupling of 48 GHz (Ref.~\cite{hepp:2014}) and the SnV$^-$ has a spin-orbit coupling of 830 GHz (Table~\ref{tab:SnVparameters}). (b) Microwave Rabi rate versus ground state strain predicted for either the SiV$^-$ or SnV$^-$, using the microwave drive field used in this work. (c) Branching ratio $\eta$ versus ground state strain. For simplicity, models in (b) and (c) neglect the orbital Zeeman effect ($f=0$), spin anisotropy ($\delta=0$) and excited state strain ($\Upsilon_e=0$). Simulations assume a background magnetic field of $B=184 \units{mT}$ oriented along the Z axis in lab coordinates (angle $\zeta=0$).
}
\label{fig:SiV_vs_SnV}
\end{center}
\end{figure}

In order to mitigate drive-induced heating, state-of-the-art SiV$^-$ experiments today operate using highly strained SiV$^-$'s, such that $\Upsilon_g \gg \lambda_g$ (see, e.g., Ref.~\cite{stas:2022}, which has a ground state splitting of 554 GHz). In this limit the ground state splitting is dominated by strain rather than spin-orbit coupling, the microwave power needed for a given Rabi rate approaches that of a free electron, and the branching ratio $\eta$ is reduced. See Fig.~\ref{fig:SiV_vs_SnV} for details.

On the other hand, because the SnV$^-$ has a much larger spin-orbit coupling it can remain coherent at 1.7 K even for $\Upsilon_g \ll \lambda_g$. In this limit the center's spin dipole moment is reduced compared to a free electron (or nitrogen-vacancy center in diamond, for example). But for the limit of moderate strain we operate at in this work ($\Upsilon_g/2\pi = 177 \units{GHz}$) the dipole moment is only reduced by a factor of a few compared to an SiV$^-$ with similar strain magnitude, Fig.~\ref{fig:SiV_vs_SnV}b. Also for a given strain magnitude the branching ratio $\eta$ is greater for the SnV$^-$, which is advantageous for readout.

In summary, the respective advantages of the SiV$^-$ versus SnV$^-$ depend on the application (e.g. the priority of minimal thermal decoherence, drive-induced heating, spin readout, emission wavelength, etc.). But the much larger spin-orbit coupling of the SnV$^-$ suppresses drive-induced heating exponentially, Eq.~\ref{eqn:T1_vs_temp_model}, while for a moderately strained center only marginally reduces the susceptibility to microwave spin control compared to that of a free electron, Fig.~\ref{fig:SiV_vs_SnV}b. Our work uses relatively high drive power because microwave delivery is not yet optimized, not because a moderately strained SnV$^-$ is that much more difficult to drive than other color centers. We therefore argue that the SnV$^-$ is a favorable choice for spin coherent quantum experiments.

\section{Coherence Model}
\label{sec:coherence_model}

\subsection{Cluster correlation expansion (CCE) calculations}

Both our measured $T_2^*$ and $T_2^\mathrm{echo}$ are shorter than would be expected if they were purely limited by the nuclear spin bath (around $1 \units{\mu s}$ and 1 ms, respectively \cite{wolfowicz:2021}). The reduction in coherence from this limit can be explained by the presence of other magnetic noise sources, mainly paramagnetic defects and other impurities in the diamond. 

\begin{figure}[tb!] 
\begin{center}
\includegraphics[width=1.0\columnwidth]{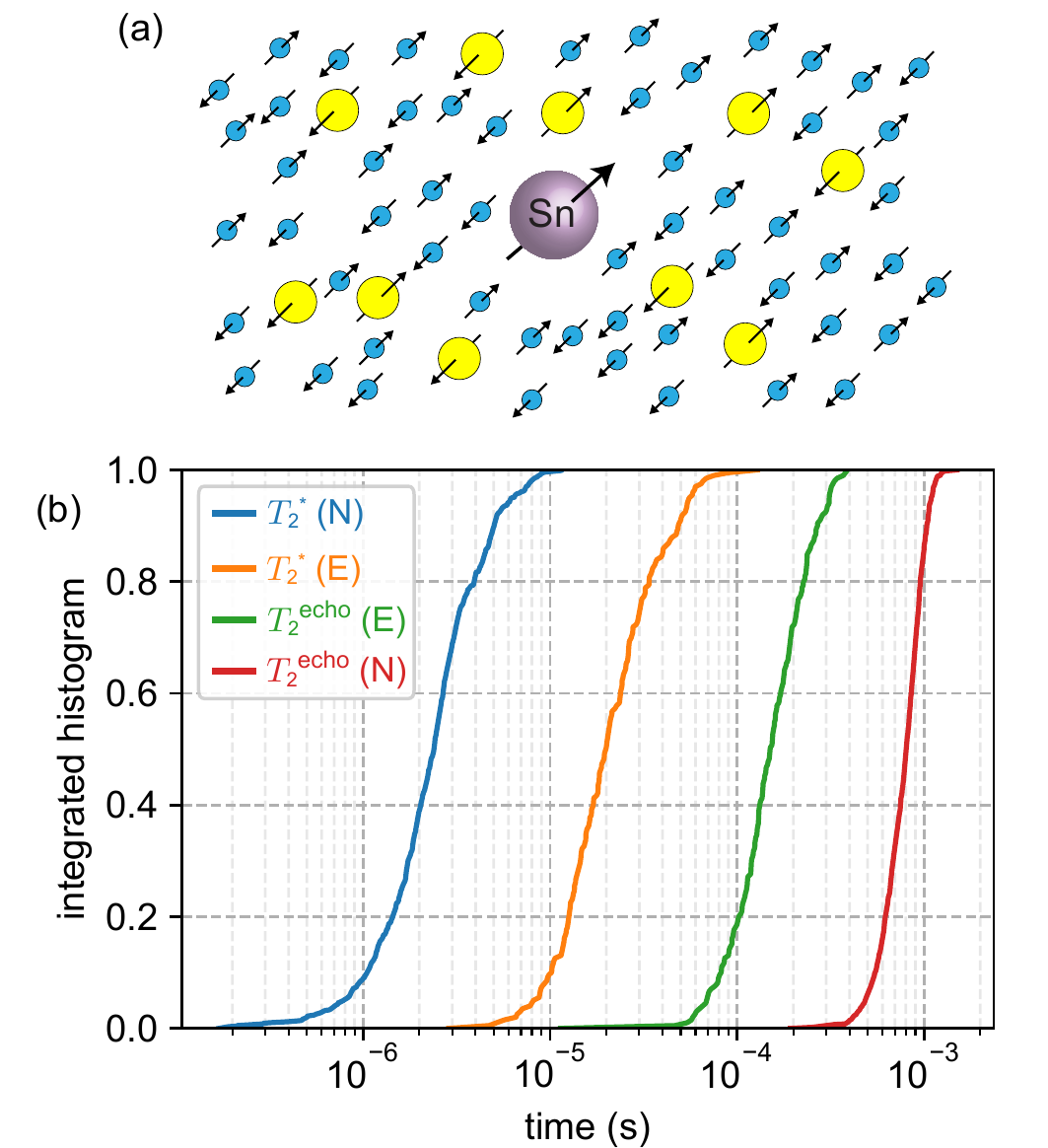}
\caption{
Coherence model. (a) The SnV$^-$ qubit is affected by surrounding baths of both nuclear spins (small circles, blue) and electron spins (larger circles, yellow). These baths interact with the SnV$^-$ qubit causing dephasing. (b) Simulated $T_2^*$ and $T_2^\mathrm{echo}$ of the qubit given dephasing due to a bath of nuclear spins (N) or electron spins (E). Lines are the cumulative distribution function of many simulations over different bath configurations. A nuclear spin bath at a concentration of natural abundance simulates the distribution of $T_2^*$ (blue line) to have a mean of $2.8 \units{\mu s}$, a median of $2.4 \units{\mu s}$ and a standard deviation of $1.8 \units{\mu s}$. The same nuclear spin bath simulates $T_2^\mathrm{echo}$ (red line) to have a mean of $795 \units{\mu s}$, median of $801 \units{\mu s}$, and a standard deviation of $190 \units{\mu s}$. An electron spin bath at concentration $8\times10^{16}$ cm$^{-3}$ simulates the distribution of $T_2^*$ (orange line) to have a mean of $25 \units{\mu s}$, a median of $20 \units{\mu s}$ and a standard deviation of $17 \units{\mu s}$. The same electronic spin bath simulates $T_2^\mathrm{echo}$ (green line) to have a mean of $165 \units{\mu s}$, median of $151 \units{\mu s}$, and a standard deviation of $74 \units{\mu s}$.
}
\label{fig:spin_bath}
\end{center}
\end{figure}

In particular, the high energy implantation used to create SnV$^-$ centers creates many nearby vacancy-related spins. We probe this contribution to decoherence by modeling our system using the cluster-correlation-expansion (CCE) technique \cite{onizhuk:2021}. In this way, the contributions of the electron and nuclear spin baths can be directly estimated. Here, we simulate up to second order (CCE-2). We find that the measured Hahn-echo time ($170 \units{\mu s}$, Fig.~\ref{fig:t2}) is consistent with a bath of $S=1/2$ electron spins that surround our qubit at a concentration of $\approx8\times10^{16} \units{cm}^{-3}$. At this concentration CCE simulations predict a Hahn-echo time of $T_2^\mathrm{echo} = 165 \pm 74 \units{\mu s}$, Fig.~\ref{fig:spin_bath} (green model, Fig.~\ref{fig:spin_bath}). 

This concentration is consistent with SRIM calculations (stopping range of ions in matter, Ref.~\cite{ziegler:2010}) calculations with the given implantation energy (370 keV) of Sn ions. These simulations estimate a concentration of Sn atoms of order $10^{16} \mathrm{cm}^{-3}$, and of vacancies produced (before annealing) of order $10^{19} \mathrm{cm}^{-3}$. Such concentration of vacancies, even with a conversion efficiency of only 1$\%$ could likely result in the observed density of spins. 

On the other hand, this electron-spin bath predicts $T_2^* = 25 \pm 17 \units{\mu s}$ (orange model, Fig.~\ref{fig:spin_bath}); longer than we observe. We explain the measured $T_2^* \approx 400 \units{ns}$ as being encompassed in the natural defect to defect variations of coherence times limited by the nuclear spin bath. For naturally abundant diamond, we simulate many random configurations of nuclear spins, where the probability of observing a $T_2^*$ of less than $1 \units{\mu s}$ (as we do in this work) is approximately 10$\%$. Our calculations assume the point-dipole approximation for the hyperfine coupling, which is known to somewhat overestimate Ramsey times \cite{onizhuk:2021}.

In addition, from the temperature dependence in Fig.~\ref{fig:t2_vs_temp}, we eliminate phonon-induced contributions to $T_2^*$, while the symmetry protection and $S=1/2$ nature makes electrical noise an unlikely contributor. We also see no appreciable change in $T_2^*$ as a function of magnetic field. Finally, similar $T_2^*$ times to what we measure are also commonly reported for the NV$^-$ center in natural diamond; e.g., Ref.~\cite{bauch:2020}. We therefore ascribe the observed $T_2^*$ to a reasonably likely configuration of nuclear spins that cause decoherence. 

In total, our measurements of coherence time in Fig.~\ref{fig:rabi}e~and~\ref{fig:t2} are consistent with an electron-spin limited $T_2^\mathrm{echo}$ and nuclear spin limited $T_2^*$. This result is similar to other experimental and computational results in related systems \cite{bourassa:2020}. We note that although the group IV centers may be insensitive to local electrical noise, they are still sensitive to magnetic noise in the spin ground state, such that careful consideration still must be made on the formation process and on the presence of nearby fabricated surfaces.

\subsection{Semiclassical model of instantaneous diffusion}

Instantaneous diffusion can be modeled semiclassically with the the characteristic dipolar coupling rate $R_\mathrm{dipolar}$ of the electron-spin bath and the central qubit:
\begin{equation}
    R_\mathrm{dipolar}=C_{B}(2\pi\gamma)(2\pi\gamma_B)\frac{\pi}{9\sqrt{3}}\mu_{0}\hbar,
    \label{eqn:semiclassical}
    \tag{C1}
\end{equation}
where $C_B$ is the concentration of bath spins and $\gamma,\gamma_B$ the gyromagnetic ratios of the qubit and bath, respectively \cite{wolfowicz:2021}. Using the dipolar coupling strength, one can estimate $T_{2}=1/R_\mathrm{dipolar}$. With a given detuning ($\delta$) of the bath from the drive frequency, the concentration of bath spins that contribute to instantaneous diffusion depends on the pulse bandwidth. The fraction of bath spins that contribute is computed by estimating the probability that the control pulse causes a spin flip, $P$. Normalizing so that $P=1$ on resonance, and assuming a Gaussian pulse shape with a bandwidth, $\sigma$ that is much larger than the spin linewidth:
\begin{equation}
    P_\mathrm{flip}=e^{-\delta^{2}/2\sigma^2}.
    \tag{C2}
\end{equation}
Here $C_B\rightarrow C_B P_\mathrm{flip}$. The assumption of a Gaussian envelope in this case is reasonable: our $\pi$-pulse time is 50 ns, but is gated by switches with a $\approx$20 ns rise time, such that a rounding of the pulse causes it to be roughly Gaussian in time (and therefore in frequency). We note that the exact shape of the pulse bandwidth is less critical to the physics at play in this experiment. 

Finally, the detuning of the qubit and bath changes as a function of magnetic field due to the difference in effective gyromagnetic ratio/$g$ factor:
\begin{equation}
    \delta=(g_B-g_0)\frac{\mu_B}{\hbar},
    \tag{C3}
\end{equation}
with $\mu_B$ the Bohr magneton. This estimate to $T_{2,ID}$ from instantaneous diffusion does not include `regular' decoherence (giving $T_{2,0}$) from this electron spin bath source. In our model we add this contribution to coherence as:
\begin{equation}
    \frac{1}{T_2}=\frac{1}{T_{2,ID}}+\frac{1}{T_{2,0}}
    \tag{C4}
\end{equation}

In this experiment, we can explain the residual coherence $T_{2,0}$ with an electron spin bath as described above. We feed this computed concentration into \ref{eqn:semiclassical}, letting us fit the observed behavior in Fig.~\ref{fig:t2_vs_b} with the only free parameters being $T_{2,0}$ (which theoretically, we know exactly from experiment) and the effective g-factor $g_B$. We note that with zero detuning, that the coherence time does not drop to zero, but instead plateaus to a finite value (here, around 25 $\mu s$) dependent on the electron-spin bath concentration.

\section{Extended Data}

\subsection{Optical characterization}

\begin{figure}[tb!] 
\begin{center}
\includegraphics[width=1.0\columnwidth]{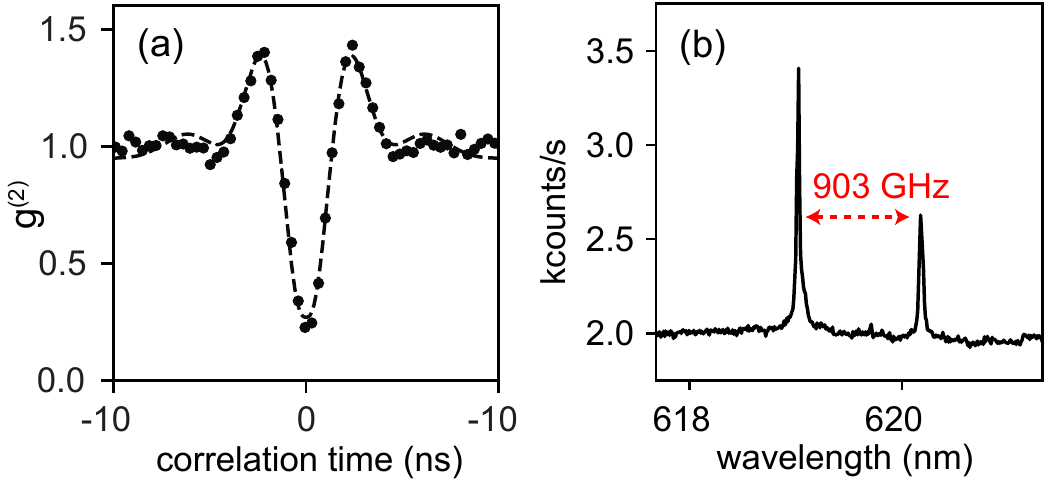}
\caption{(a) Optical on-resonance $\mathrm{g}^{(2)}$ measurement. The dip below one half of the background counts demonstrates there is a single emitter within the confocal spot at which the measurements in this work were taken. The oscillatory behavior and subsequent fit suggests optical Rabi oscillations. (b) Photoluminescence spectrum at 1.7 K and zero magnetic field. The ground state splitting is $\Delta_g = 903.0 \pm 0.7 \units{GHz}$, obtained from a two Lorentzian fit.}
\label{fig:pl}
\end{center}
\end{figure}

In Fig.~\ref{fig:pl}a we show a photoluminescence (PL) spectrum of this emitter, taken under above resonant excitation (2 mW of 532 nm light), at 1.7 K, and at zero magnetic field. The lower (higher) wavelength peaks are the transitions between the degenerate $\ket{A}$ and $\ket{B}$ spin states to the $\ket{1}$ and $\ket{2}$ ($\ket{3}$ and $\ket{4}$) spin states. These transitions are measured at 619.0263 and 620.1757 nm, respectively. A two Lorentzian fit gives a splitting of $\Delta_g/2\pi = 903 \pm 0.7~\unit{GHz}$. 

On the same confocal spot we also do a $\mathrm{g}^{(2)}$ correlation measurement, Fig.~\ref{fig:pl}. The measurement is taken using both continuous wave above-resonant excitation and continuous wave resonant excitation. Collection is on the phonon sideband. The dip below $1/2$ of the background count rate confirms we are measuring a single-photon emitter. Background signal above zero is likely due to PL from nearby emitters.

\begin{figure}[tb!] 
\begin{center}
\includegraphics[width=1.0\columnwidth]{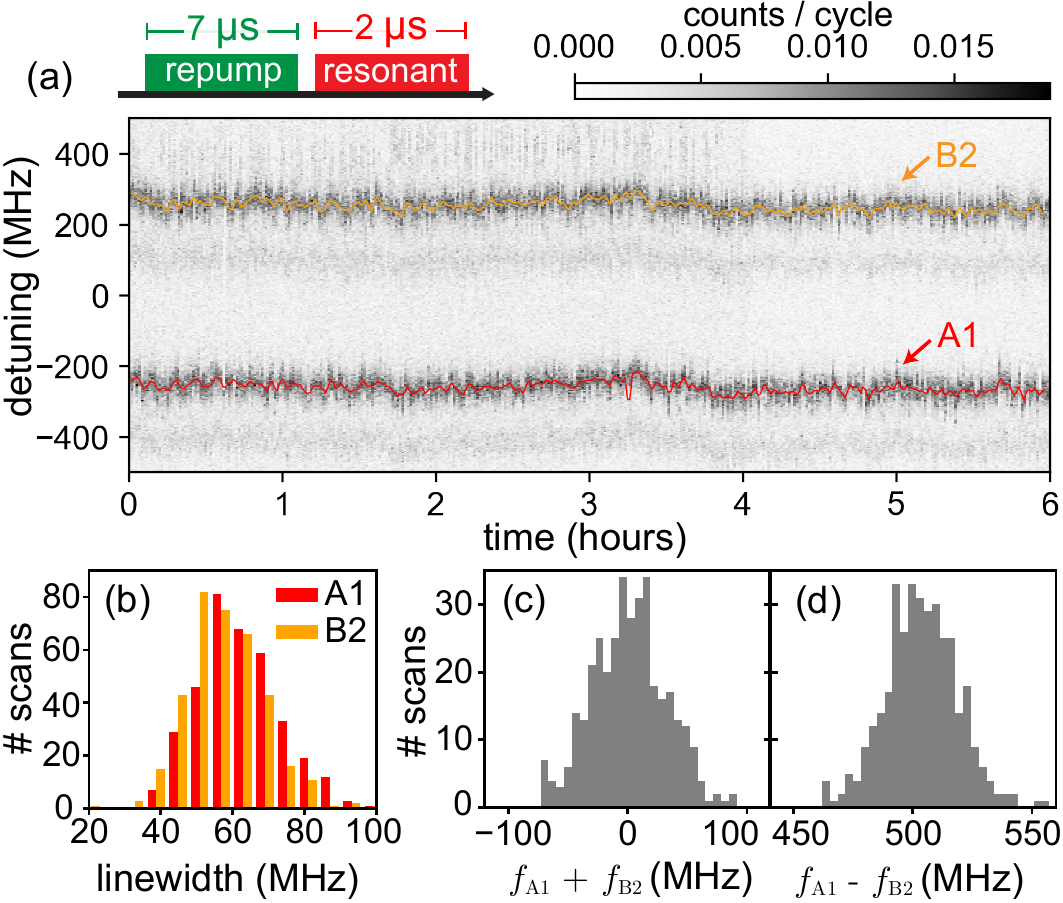}
\caption{Optical stability and linewidth. (a) Photoluminescence excitation (PLE) at field $|\Vec{B}| = 150 \units{mT}$ and angle $\zeta = 110^{\degree}$. The spin preserving (A1 and B2) transitions are resolved with a linewidth of $60 \pm 10 \units{MHz}$ each. (c) Sum and (d) difference of the transition frequencies.
}
\label{fig:stability}
\end{center}
\end{figure}

Next, we show a measurement of photoluminescence excitation (PLE) taken over 6 h at a scan rate of 20 MHz/s, Fig.~\ref{fig:stability}a. 
This measurement is done at the same $\Vec{B}$ used in the main text spin control data. The measured A1 and B2 transition linewidths are $\kappa_\mathrm{A1}/2\pi = 62.7 \pm 11.8 \units{MHz}$ and $\kappa_\mathrm{B2}/2\pi = 57.7 \pm 11.6 \units{MHz}$, respectively, Fig.~\ref{fig:stability}b. The center frequency of these transitions wanders by a linewidth or less over hour timescales: the common mode shift $f_\mathrm{A1} + f_\mathrm{B2}$ of these transitions has a standard deviation of $30.0 \units{MHz}$, Fig.~\ref{fig:stability}c. The difference $|f_\mathrm{A1} - f_\mathrm{B2}|$ between these transitions is $504.0 \pm 16.5 \units{MHz}$, Fig.~\ref{fig:stability}d.

\begin{figure}[htb] 
\begin{center}
\includegraphics[width=1.0\columnwidth]{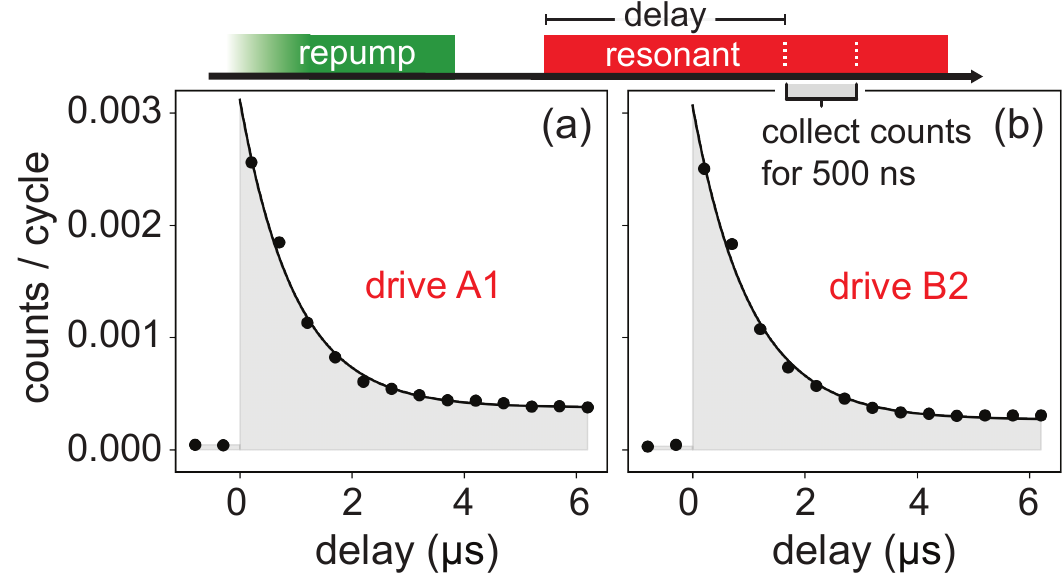}
\caption{Spin selective initialization by driving the (a) A1 and (b) B2 transition. Measured steady-state bound initialization fidelity to be $\gtrsim 90\%$. Most background counts are likely due to scatter of the excitation laser into the collection path; the actual initialization fidelity is closer to unity. The same resonant drive is used for spin selective readout.
}
\label{fig:init}
\end{center}
\end{figure}

\subsection{Contributions to pulse infidelity}
\label{sec:infidelity}

\begin{figure}[htb] 
\begin{center}
\includegraphics[width=1.0\columnwidth]{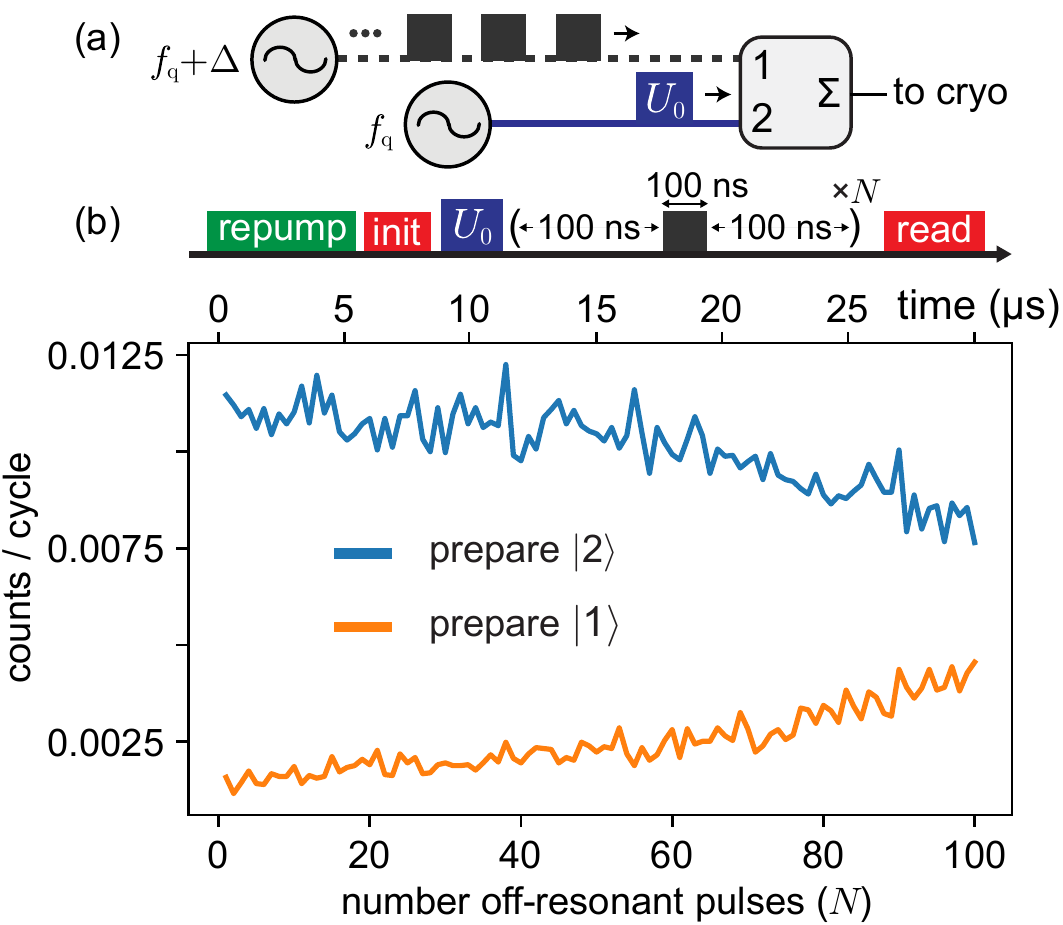}
\caption{
The qubit is prepared in either the $\ket{2}$ or $\ket{1}$ state using a pulse $U_0 = \pi$ or $U_0 = I$, respectively. Then a series of $N$ off-resonant pulses is applied. These pulses are the same power as the $\pi$ pulse, but are detuned by $100 \units{MHz}$ from the qubit frequency $\omega_q/2\pi = 3.9349 \units{GHz}$. (b) Qubit readout contrast decays with increased $N$, presumably due to heating caused by the off-resonant pulses.
}
\label{fig:N_offres}
\end{center}
\end{figure}

\begin{figure}[htb] 
\begin{center}
\includegraphics[width=1.0\columnwidth]{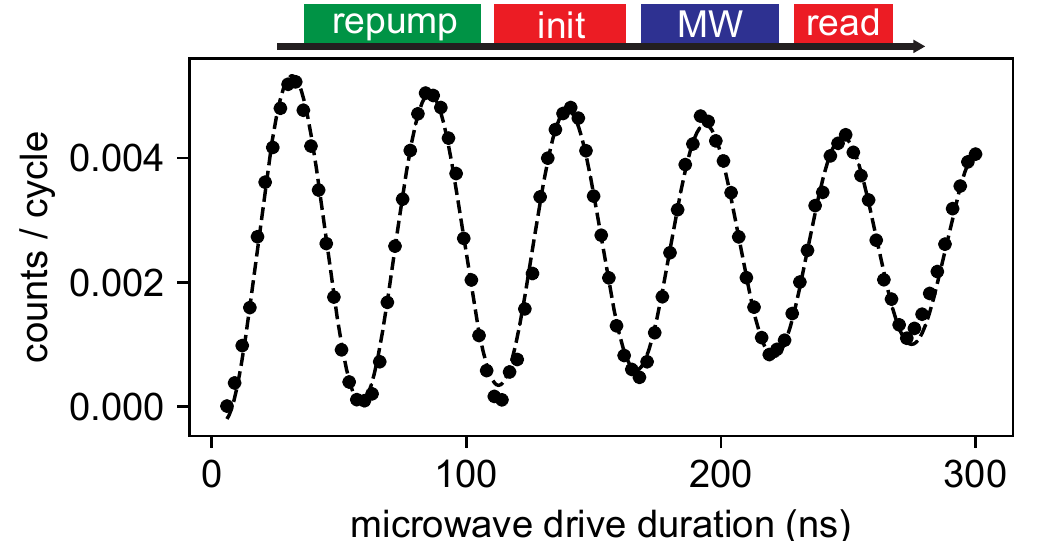}
\caption{Rabi oscillations at $\Omega_\mathrm{MW}/2\pi = 36.98 \pm 0.03 \units{MHz}$. Drive-induced heating may cause contrast reduction, even though the sample temperature rises to only 1.74 K (from 1.70 K) during this measurement.}
\label{fig:rabi_vs_power}
\end{center}
\end{figure}

Drive-induced heating is another source of pulse infidelity and is the main drawback of microwave spin control. We study drive-induced heating with the following experiment. First, we alternatively prepare the qubit in either $\ket{2}$ or $\ket{1}$ by applying either a $\pi$ pulse or the identity $I$, respectively. Next, we apply a series of $N$ off-resonant pulses which have nominally the same power and duration as the $\pi$ pulse but are detuned from the qubit frequency by 100 MHz (much more than the $\approx3 \units{MHz}$ qubit linewidth). For the data in Fig.~\ref{fig:N_offres}, these pulses are 100 ns in duration with a buffer of 200 ns between each. Upon application of these $N$ pulses the qubit state gradually decays toward the maximally mixed state, indicating the pulses are heating the qubit causing $T_1$-like decay. This effect is qualitatively similar to the extra infidelity measured in the data in Fig.~\ref{fig:rb} and the decay observed at high Rabi rates, Fig.~\ref{fig:rabi_vs_power}.

\section{Optical qubit control}
\label{sec:optical_spin_control}
An alternative approach to coherent spin control is optical Raman driving \cite{debroux:2021}. Optical Raman driving has the advantage of minimal drive-induced heating compared to the microwave spin control presented in this work. It has the disadvantage, however, of drive-induced dephasing due to weak excitation of an optical transition. We argue that microwave driving is a clearer path toward high-fidelity spin control of SnV$^-$'s. To quantify this argument, we briefly overview optical Raman driving, then present an experimental demonstration. 

An optical Raman drive consists of two laser tones, detuned from each other by the qubit frequency and also each detuned by $\Delta_\mathrm{Raman}$ from the optical transitions that separate the qubit states from the excited state (e.g. the A1 and A2 transitions of an SnV$^-$), Fig.~\ref{fig:raman}a. These tones create a beat note at the qubit frequency, coherently shuttling population back and forth at the Rabi rate \cite{debroux:2021}: $\Omega_\mathrm{Raman} = s \kappa^2 / (\sqrt{\eta} 4 \Delta_\mathrm{Raman})$.

Here, $s=p/p_\mathrm{sat}$ is the optical drive power normalized by the saturation power, $\kappa$ is the linewidth of the optical transitions, and $\eta$ is the branching ratio. The Rabi rate is greatest at smaller detunings $\Delta_\mathrm{Raman}$, where the laser interacts more strongly with the atom. Driving \textit{on} a transition, however, initializes and measures the qubit---an action which destroys phase coherence. By the same mechanism the Raman drive also causes qubit dephasing at rate \cite{debroux:2021} $\Gamma_\mathrm{Raman} = s \kappa^3 / (8 \Delta_\mathrm{Raman}^2)$.

Raman driving is therefore effective only when the rate of coherent rotation is much greater than that of dephasing, such that $\Omega_\mathrm{Raman} / \Gamma_\mathrm{Raman} = 2 \Delta_\mathrm{Raman} / \sqrt{\eta} \kappa \gg 1$, Fig.~\ref{fig:raman}b. This means that Raman driving works best when the laser detuning is large compared to a linewidth. Because the SnV$^-$ has a transform limited linewidth of $\kappa/2\pi \gtrsim 30 \units{MHz}$ \cite{iwasaki:2017,trusheim:2020,gorlitz:2020}, detunings of many gigahertz or more are required for high-fidelity optical Raman gates. 

\subsection{Coherent population trapping (CPT)}
\label{sec:cpt}
Coherent population trapping (CPT), Fig.~\ref{fig:ple}b, is a useful first step toward spin control. In CPT, population is shuttled between qubit states $\ket{1}$ and $\ket{2}$ via transitions to a third state at much higher energy $\ket{A}$. Under a Raman drive, a $\lambda$ system in the $\{ \ket{1}, \ket{2}, \ket{A} \}$ basis has the Hamiltonian \cite{debroux:2021}
\begin{equation}
\hat{H}_\mathrm{CPT}\:=\:\begin{pmatrix} 
\delta _{1} & 0 & \frac{1}{2} \Omega_\mathrm{A1}\\
0 & \delta _{2} & \frac{1}{2} \Omega_\mathrm{A2}\\
\frac{1}{2} \Omega_\mathrm{A1} & \frac{1}{2} \Omega_\mathrm{A2} & 0
\end{pmatrix}, \\
\label{eqn:cpt}
\tag{E1}
\end{equation}
where $\delta_1$, $\delta_2$ are detunings from the eigenenergies of $\ket{1}$ and $\ket{2}$, respectively, and $\Omega_\mathrm{A1}$ and $\Omega_\mathrm{A2}$ are coupling rates that increase with Raman drive strength. 

In the absence of a Raman drive, $\Omega_\mathrm{A1} = \Omega_\mathrm{A2} = 0$ and $\hat{H}_\mathrm{CPT}$ is diagonal. When the Raman drive is on and detuning approaches zero, the basis in which Eq.~\ref{eqn:cpt} is diagonal has an eigenstate that is a linear combination of the $\ket{1}$ and $\ket{2}$ qubit states \textit{only} and is hence ``dark'', having no contribution in the excited state and emitting no photons. For nonzero detuning the eigenstates of the system change such that all have contributions in the $\ket{A}$ state, leading to a bright peak on either side of the dip.

These dynamics are seen in Fig.~\ref{fig:ple}b, where there is a measured dip at the spin frequency surrounded by two bright peaks. This measurement is compared to a numerical model (dashed line) with parameters: $\Omega_\mathrm{A1}/2\pi = 8.95 \units{MHz}$, $\Omega_\mathrm{A2}/2\pi = 0.42 \units{MHz}$, and optical and spin linewidths of $30$ and $3 \units{MHz}$, respectively.

\subsection{Rabi oscillations with optical control}
\begin{figure}[htb] 
\begin{center}
\includegraphics[width=1.0\columnwidth]{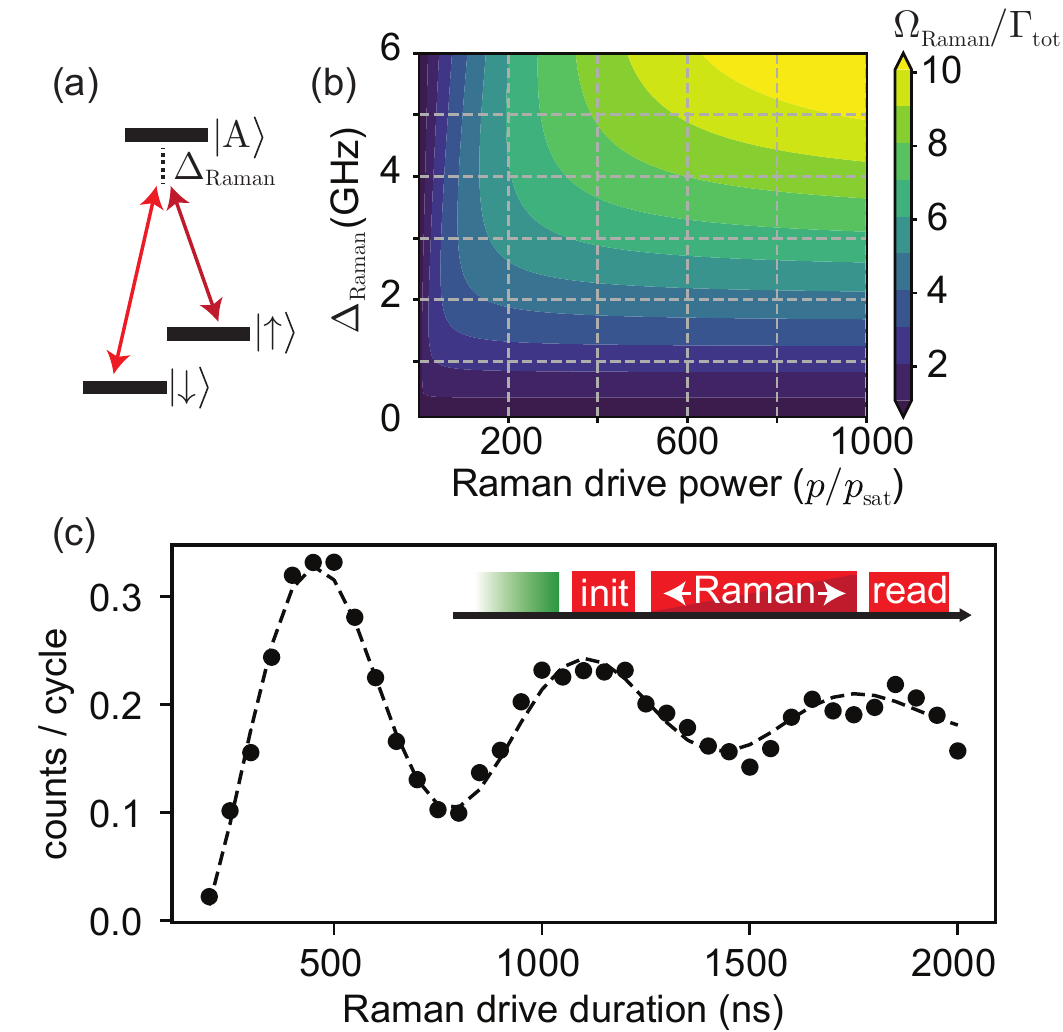}
\caption{
Optical spin control. (a) A $\lambda$ system is driven by a two-tone drive, with tones detuned by $\Delta_\mathrm{Raman}$ from the optical transitions. (b) Modeled Rabi rate $\Omega_\mathrm{Raman}$, divided by the total dephasing rate $\Gamma_\mathrm{tot} = \Gamma_\mathrm{Raman} + \Gamma^*$.  Here, $\Gamma_\mathrm{Raman}$ is extra dephasing introduced by the optical Raman drive and $\Gamma^*$ is the dephasing rate, simulated here to $\Gamma^*/2\pi = 1 \units{MHz}$. Optical Raman gates work better ($\Omega_\mathrm{Raman} / \Gamma_\mathrm{tot} \gg 1$) at high drive powers and large detunings. (c) Measured Rabi oscillations at a rate of $\Omega_\mathrm{Raman}/2\pi = 2.94 \units{MHz}$ and a dephasing rate of $\Gamma_\mathrm{tot}/2\pi = 0.74 \units{MHz}$. Qubit frequency is 4.845 GHz.
}
\label{fig:raman}
\end{center}
\end{figure}

After calibrating the Raman drive frequency and power using coherent population trapping, we measure Rabi oscillations, Fig.~\ref{fig:raman}c. For this data, the laser is red detuned by 1.2 GHz from the A1 transition and 2 GHz from the B2 transition. The observed gate fidelity is much lower than our demonstrated microwave Rabi oscillations, Fig.~\ref{fig:rabi}, but could be improved by going to larger detunings and higher drive powers.

In conclusion, microwave driving outperformed optical Raman driving in this experimental setup. The main disadvantage of optical spin control is drive-induced dephasing. Because of the SnV$^-$'s transform limited linewdith of $\kappa/2\pi \gtrsim 30 \units{MHz}$ \cite{iwasaki:2017,trusheim:2020,gorlitz:2020} and the requirement that $\Delta_\mathrm{Raman}/\kappa \gg 1$ for high-fidelity control, this technique requires large $\Delta_\mathrm{Raman}$, which can present a technical challenge. Optical Raman driving may still be useful in the SnV$^-$ platform, however, for applications where microwave power delivery is difficult.

\section{Sample preparation}
\label{sec:sample_prep}
The diamond sample used in this work is a $2 \units{mm} \times 2 \units{mm} \times 0.5 \units{mm}$ electronic grade plate, with $\langle100\rangle$ face orientation (Element Six). The sample is patterned with an array of nanopillars to enhance collection efficiency, Fig.~\ref{fig:packaging}d, similar to the device in Ref.~\cite{rugar:2019}. The sample was first processed in a triacid mixture to remove surface contaminants and any graphitization, and then etched in $\mathrm{O}_2$ plasma to remove ~$1 \units{\mu m}$ of surface polishing damage. SnV$^-$'s are then introduced by commercially implanted (CuttingEdge Ions) Sn$^-$ ions at a dose of $2\times10^{11} \units{cm}^{-2}$ with an energy of $370 \units{keV}$ leading to an implant depth of 92 nm with 18 nm straggle. After implantation, the sample is sequentially annealed at 800$\degree$C for 30 min and 1100$\degree$C for 90 min. Both anneals were performed at pressures $<10^{-4}$ torr. 200 nm of Si$_{x}$N$_{y}$ is then deposited with a chemical vapor deposition tool to serve as a hard mask. The nanopillar array is defined lithographically in in hydrogen silsesquioxane FOx-16 with a 100 keV electron beam writer. The nitride hard mask is first anisotropically etched through with a SF$_{6}$, CH$_{4}$, and N$_{2}$ chemistry, and the exposed diamond etched 500 nm with an O$_{2}$ plasma. Resist and mask layers are lastly stripped by soaking the sample in hydrofluoric acid for 20 min.

\section{Experimental details}

\subsection{Setup}
\label{sec:setup}

\begin{figure}[tb!] 
\begin{center}
\includegraphics[width=1.0\columnwidth]{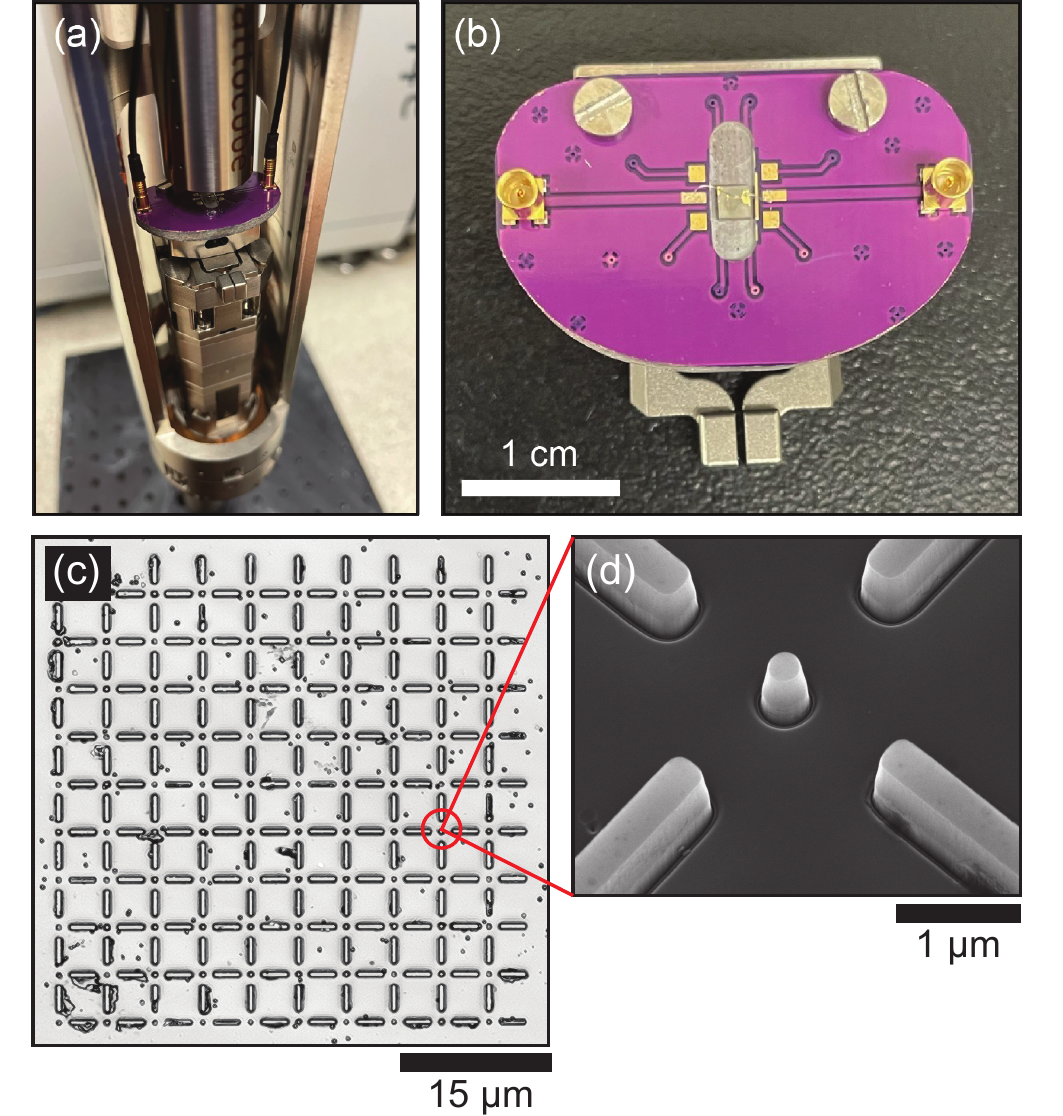}
\caption{
(a) Photograph of the sample space. Microwaves are routed through cables (thin black lines) that connect to additional microwave cables higher up in the cryostat. (b) Printed circuit board. Coplanar waveguides on the circuit board connect to surface mount soldered connectors on the left and right of the board, and to bonding pads closer to the center. The diamond chip is placed at the center of the circuit board, in a groove milled to be of similar depth as the chip height. (c) Optical image of an array of nanopillars. An example nanopillar is circled in red. 
(d) Scanning electron microscope image of a
single nanopillar.
}
\label{fig:packaging}
\end{center}
\end{figure}

The device in this work is measured in a $^4$He bath cryostat (attoDRY2100, Attocube) with a two-axis vector magnet, a rotation stage, and a base temperature of 1.7 K. Optical excitation and collection are done using a home-built confocal microscope, including a light-emitting diode (LED) used to illuminate the sample during alignment. A cryogenic objective (LT-APO/VISIR/0.82, Attocube) is used to focus light onto the sample.

Resonant excitation near 619 nm, red, is done using a Ti:sapphire laser tuned near 906 nm and mixed with $2 \units{\mu m}$ light (M-Squared). Off-resonant repump excitation is done using a 532 nm green laser (pickoff from a Verdi V-10, Coherent). Green and red tones are gated using AOMs (acousto-optic modulators: 4C2C-532-AOM and 4C2C-633-AOM, Gooch and Housego). For some data (e.g., photoluminescence excitation scans and coherent population trapping data), transitions are excited using the first sideband created by an EOM (electro-optic modulator: PM-0S5-PFU-PFU-620, Eospace). Excitation paths are filtered before entering the cryostat in order to remove unwanted excitation light such as fiber photoluminescence. The collection path is filtered to collect the SnV$^-$ phonon sideband.

Microwave control is done using a vector signal generator (SG396, Stanford Research Systems), controlled using an AWG (arbitrary waveform generator: Pulse Streamer 8/2, Swabian). Data are collected using single-photon avalanche diodes. Microwave switches (ZASWA-2-50DRA+, rise time 20 ns, Mini-Circuits) are used to gate the connection between photon counters and DAQ (data acquisition: PCIe-6321, National Instruments), and to gate signal generator outputs.
The microwave signal is amplified before entering the cryostat using a high-power amplifier (50S1G4AM2, Amplifier Research). Inside the cryostat, two 0 dB cryo attenuators (2082-6418-dB-CRYO, XMA) each are placed on the microwave input and output lines, in order to better thermalize their center conductors by creating metallic contact between the center and outer conductors. The output microwave line is connected to a high-power 50 Ohm termination.

Inside the cryostat, the microwave input-output lines connect to a printed circuit board (PCB), Fig.~\ref{fig:packaging}, using surface mount soldered launchers whose center conductors feed into bonding pads. On chip there is also one bonding pad ($300 \units{nm}$ gold, with $15 \units{nm}$ of titanium underneath for adhesion), left over from a failed attempt to fabricate an on-chip microwave bias line. Gold lift-off failed due to uneven resist spinning on the small chip. Several wire bonds connect one PCB bonding pad to this on-chip pad, and then one long wire bond connects this on-chip pad to the other PCB pad, draped to be as close as possible to the measured SnV$^-$. This unusual assembly made it easier to position the draped wire bond close to the measured SnV$^-$ (compared to instead using one wire bond connecting both PCB bonding pads, only). However, this configuration creates contact between the microwave bias line and diamond chip at the gold pad, which is $\approx 1 \units{mm}$ away from the measured SnV$^-$. This may result in more heating than necessary. 

\begin{figure*}[htb] 
\begin{center}
\includegraphics[width=2.0\columnwidth]{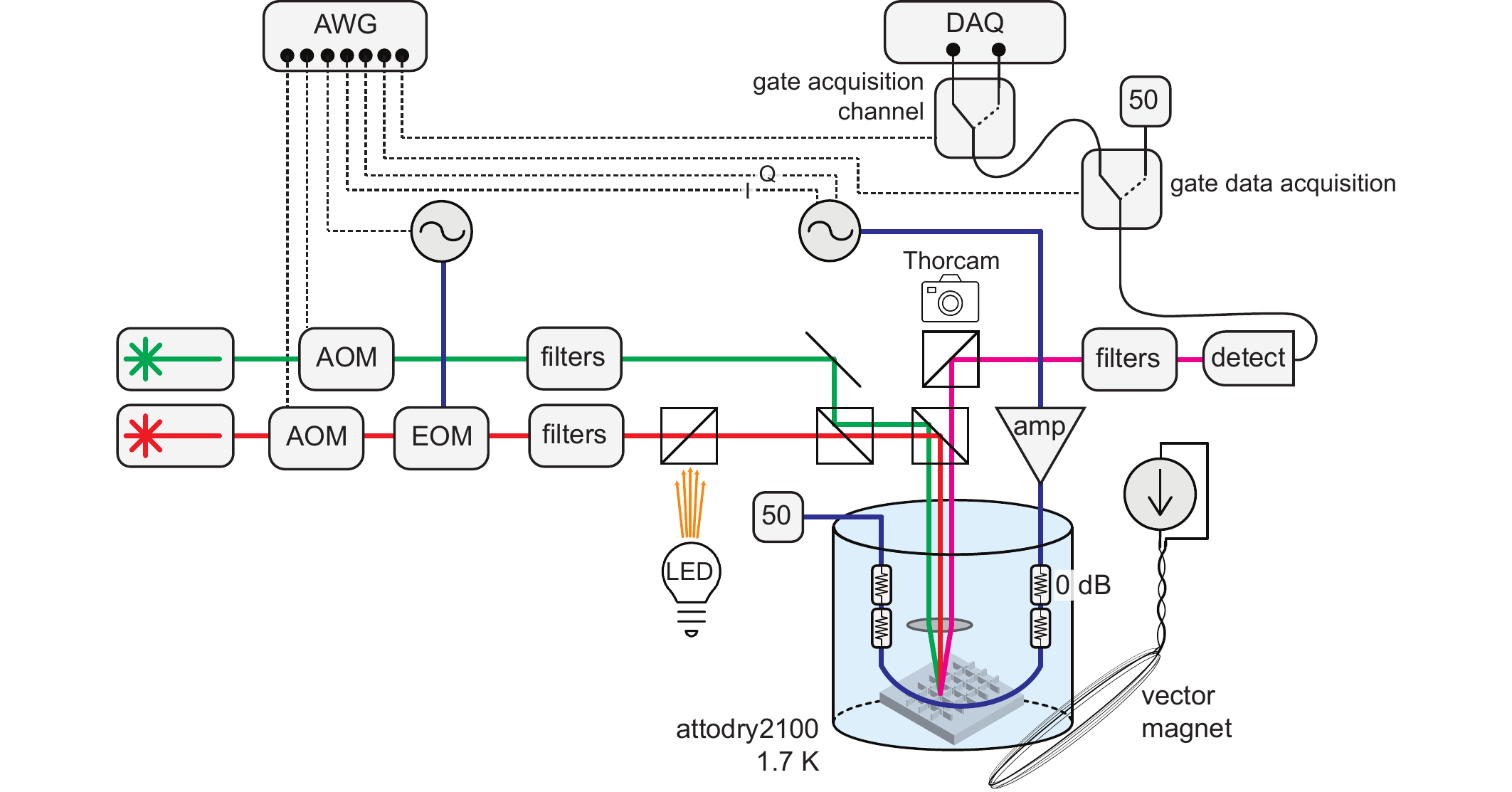}
\caption{
Experimental schematic.
}
\label{fig:schematic}
\end{center}
\end{figure*}

See Fig.~\ref{fig:schematic} for a full experimental schematic.

\vspace{-\baselineskip}

\subsection{Microwave power characterization}
\label{sec:mw_power_characterization}
In this section we characterize the microwave power used to control our SnV$^-$ qubit.

For the operating conditions in Fig.~\ref{fig:rabi} (Rabi rate $\approx 21 \units{MHz}$ at $\omega_q/2\pi = 3.836 \units{GHz}$) we use 15.7 dBm of continuous power output from our microwave signal generator. Using a calibrated vector network analyzer (VNA) we measure 19.4 dB of attenuation between the signal generator output to the amplifier input. This includes extra attenuation added to limit maximum power delivery. Using a calibrated VNA at low power we measure our microwave amplifier's gain to be 53.0 dB. This suggests 49.3 dBm of output from the microwave amplifier; too high for us to measure directly. Because of the specified power threshold for gain compression, however, we expect to have the slightly reduced power of $\approx 48 \units{dBm}$ at the amplifier output.

To determine the current running through the wire bond driving our qubit we now characterize microwave loss after the amplifier output. First, we measure -13.88 dB of transmission between the amplifier output and termination on the output line of the cryostat. Since the cable and packaging is nearly symmetrical we therefore assume $\approx 7 \units{dB}$ of loss/reflection between the amplifier output and wire bond midpoint, which is adjacent to the qubit. We therefore estimate $\approx 41 \units{dBm}$ of microwave power running through the wire bond, corresponding to a microwave current of $\approx 0.5 \units{A}$.

\subsection{List of data}
\label{sec:operating_conditions}

In Table~\ref{tab:operating_conditions} we provide a summary of data and operating conditions.

\begin{table*}[htb]
\caption{Summary of data and operating conditions. Excitation (red) and repump (green) powers are specified in continuous-wave (cw) and measured immediately before the cryostat window. Note that powers are most helpful as an order-of-magnitude reference: the degree to which the laser pulse interacts with the SnV$^-$ depends on polarization and alignment conditions. For some measurements the laser carrier wavelength is detuned from resonance, and transitions are driven by a sideband created by an electro-optic modulator. Durations labeled cw indicate the drive is always on. Pulse sequences have $\approx 1 \units{\mu s}$ buffer steps placed between different segments of the pulse sequence, and delays with all instruments off before starting the next sequence. Cycle time is the duration of each single-shot experiment. In some experiments the timing is changed throughout the data set (e.g. in $T_1$ versus temperature, cycle time is shortened as temperature increases). In these cases the longest duration is reported.}
  \begin{center}
    \begin{tabular}{ |p{2.4cm}|p{1.5cm}|p{1.5cm}|p{1.8cm}|p{1.8cm}|p{1.8cm}|p{1.4cm}|p{1.4cm}|p{1.4cm}|p{1.4cm}|  }
         \hline
         Data & Figure & magnet: \mbox{$|\Vec{B}|$} & magnet: \mbox{angle $\zeta$} & Excitation (red) pow. & Repump (green) pow. & Repump duration & Init. duration & Readout duration & Cycle time  \\
         \hline
         \hline
         $\omega_q/2\pi$ vs. $\zeta$ & Fig.~\ref{fig:ple}a & 184 mT & swept & varied & $180 \units{\mu W}$ & $25 \units{\mu s}$ &  & $5 \units{\mu s}$ & $33 \units{\mu s}$ \\
         A1/B2 vs. $\zeta$ & Fig.~\ref{fig:ple}a & 184 mT & swept  & not recorded & $180 \units{\mu W}$ & $7 \units{\mu s}$ &  & $3 \units{\mu s}$ & $13 \units{\mu s}$ \\
         CPT & Fig.~\ref{fig:ple}b & 150 mT & $110^{\degree}$ & $0.64 \units{\mu W}$ & not recorded & $10 \units{\mu s}$ &  & $10 \units{\mu s}$ & $30 \units{\mu s}$ \\
         Zeeman effect & Fig.~\ref{fig:ple}c & swept & $83^{\degree}, 110^{\degree}$ & $7 \units{\mu W}$ & $160 \units{\mu W}$ & $7 \units{\mu s}$ &  & $2 \units{\mu s}$ & $12 \units{\mu s}$ \\
         ODMR & Fig.~\ref{fig:rabi}b & 150 mT & $110^{\degree}$ & $0.13 \units{\mu W}$ & $170 \units{\mu W}$ & $12 \units{\mu s}$ & $6 \units{\mu s}$ & $2 \units{\mu s}$ & $32 \units{\mu s}$ \\
         Rabi (1d) & Fig.~\ref{fig:rabi}c & 150 mT & $110^{\degree}$ & $4 \units{\mu W}$ & $190 \units{\mu W}$ & $12 \units{\mu s}$ & $9 \units{\mu s}$ & $4 \units{\mu s}$ & $35 \units{\mu s}$ \\
         Rabi (2d) & Fig.~\ref{fig:rabi}c & 150 mT & $110^{\degree}$ & $4 \units{\mu W}$ & $190 \units{\mu W}$ & $12 \units{\mu s}$ & $9 \units{\mu s}$ & $4 \units{\mu s}$ & $35 \units{\mu s}$ \\
         Rabi vs. amp. & Fig.~\ref{fig:rabi}d & 150 mT & $110^{\degree}$ & $1.0 \units{\mu W}$ & $190 \units{\mu W}$ & $50 \units{\mu s}$ & $8 \units{\mu s}$ & $3 \units{\mu s}$ & $71 \units{\mu s}$ \\
         Ramsey (1d) & Fig.~\ref{fig:rabi}e & 150 mT & $110^{\degree}$ & $4 \units{\mu W}$ & $190 \units{\mu W}$ & $12 \units{\mu s}$ & $9 \units{\mu s}$ & $3 \units{\mu s}$ & $88 \units{\mu s}$ \\
         Ramsey (2d) & Fig.~\ref{fig:rabi}e & 150 mT & $110^{\degree}$ & $4 \units{\mu W}$ & $190 \units{\mu W}$ & $30 \units{\mu s}$ & $9 \units{\mu s}$ & $3 \units{\mu s}$ & 0.124 ms \\ 
         $\pi$-pulse fidelity & Fig.~\ref{fig:rb} & 150 mT & $110^{\degree}$ & $4 \units{\mu W}$ & $190 \units{\mu W}$ & $300 \units{\mu s}$ & $12 \units{\mu s}$ & $4 \units{\mu s}$ & 0.672 ms \\   
         Clifford fidelity & Fig.~\ref{fig:rb} & 150 mT & $110^{\degree}$ & $4 \units{\mu W}$ & $190 \units{\mu W}$ & $32 \units{\mu s}$ & $9 \units{\mu s}$ & $3 \units{\mu s}$ & 0.127 ms \\
         Hahn-echo & Fig.~\ref{fig:t2} & 150 mT & $110^{\degree}$ & $4 \units{\mu W}$ & $190 \units{\mu W}$ & $32 \units{\mu s}$ & $12 \units{\mu s}$ & $4 \units{\mu s}$ & 0.916 ms \\   
         CPMG2 & Fig.~\ref{fig:t2} & 150 mT & $110^{\degree}$ & $4 \units{\mu W}$ & $190 \units{\mu W}$ & $32 \units{\mu s}$ & $12 \units{\mu s}$ & $4 \units{\mu s}$ & 1.227 ms \\   
         XY4 & Fig.~\ref{fig:t2} & 150 mT & $110^{\degree}$ & $4 \units{\mu W}$ & $190 \units{\mu W}$ & $64 \units{\mu s}$ & $12 \units{\mu s}$ & $4 \units{\mu s}$ & 1.491 ms \\   
         XY8 & Fig.~\ref{fig:t2} & 150 mT & $110^{\degree}$ & $4 \units{\mu W}$ & $190 \units{\mu W}$ & $128 \units{\mu s}$ & $12 \units{\mu s}$ & $4 \units{\mu s}$ & 2.130 ms \\   
         XY16 & Fig.~\ref{fig:t2} & 150 mT & $110^{\degree}$ & $4 \units{\mu W}$ & $190 \units{\mu W}$ & $128 \units{\mu s}$ & $12 \units{\mu s}$ & $4 \units{\mu s}$ & 2.537 ms \\      
         $T_2^\mathrm{echo}$ vs. $|\Vec{B}|$ & Fig.~\ref{fig:t2_vs_b} & swept & $110^{\degree}$ & varied & $190 \units{\mu W}$ & $32 \units{\mu s}$ & $12 \units{\mu s}$ & $4 \units{\mu s}$ & 0.916 ms \\      
         $T_1$ vs. temp & Fig.~\ref{fig:t2_vs_temp} & 150 mT & $110^{\degree}$ & $4 \units{\mu W}$ & $200 \units{\mu W}$ & $132 \units{\mu s}$ & $9 \units{\mu s}$ & $3 \units{\mu s}$ & 30.3 ms \\      
         $T_2^\mathrm{echo}$ vs. temp & Fig.~\ref{fig:t2_vs_temp} & 150 mT & $110^{\degree}$ & $4 \units{\mu W}$ & $200 \units{\mu W}$ & $32 \units{\mu s}$ & $9 \units{\mu s}$ & $3 \units{\mu s}$ & 1.110 ms \\    
         $T_2^*$ vs. temp & Fig.~\ref{fig:t2_vs_temp} & 150 mT & $110^{\degree}$ & $4 \units{\mu W}$ & $200 \units{\mu W}$ & $18 \units{\mu s}$ & $9 \units{\mu s}$ & $3 \units{\mu s}$ & 0.100 ms \\    
         $\mathrm{g}^{(2)}$ & Fig.~\ref{fig:pl}a & 0 mT &  & $14 \units{\mu W}$ & $35 \units{\mu W}$ & cw &  & cw &  \\
         PL & Fig.~\ref{fig:pl}b & 0 mT &  &  & $2 \units{mW}$ & cw &  & cw &  \\
         PLE vs. time & Fig.~\ref{fig:stability} & 150 mT & $110^{\degree}$ & $7 \units{\mu W}$ & $160 \units{\mu W}$ & $7 \units{\mu s}$ &  & $2 \units{\mu s}$ & $12 \units{\mu s}$ \\
         spin init. & Fig.~\ref{fig:init} & 150 mT & $110^{\degree}$ & $7 \units{\mu W}$ & $160 \units{\mu W}$ & $160 \units{\mu s}$ &  & $8 \units{\mu s}$ & $39 \units{\mu s}$ \\
         off res. MW & Fig.~\ref{fig:N_offres} & 150 mT & $110^{\degree}$ & $4 \units{\mu W}$ & $190 \units{\mu W}$ & $264 \units{\mu s}$ & $9 \units{\mu s}$ & $3 \units{\mu s}$ & 0.823 ms \\
         faster Rabi & Fig.~\ref{fig:rabi_vs_power} & 80 mT & $110^{\degree}$ & $13 \units{\mu W}$ & $100 \units{\mu W}$ & $50 \units{\mu s}$ & $8 \units{\mu s}$ & $3 \units{\mu s}$ & $71 \units{\mu s}$ \\
         Raman control & Fig.~\ref{fig:raman}b & 175 mT & $122^{\degree}$ & $67 \units{\mu W}$ & not recorded & $6 \units{\mu s}$ & $2 \units{\mu s}$ & $2 \units{\mu s}$ & $33 \units{\mu s}$ \\
         \hline
    \end{tabular}
  \label{tab:operating_conditions}
  \end{center}
\end{table*}

\bibliography{main}

\end{document}